\documentclass[useAMS,usenatbib,usegraphicx]{mn2e}
\usepackage{amsmath, amssymb}
\usepackage{bm}

\newcommand{\bfssA}{\mbox{\boldmath $\mathsf{A}$}}
\newcommand{\bfssO}{\mbox{\boldmath $\mathsf{O}$}}
\newcommand{\bfssLambda}{\mbox{\boldmath $\mathsf{\Lambda}$}}
\newcommand{\cross}{\mbox{\boldmath $\times$}}
\newcommand{\bcdot}{\mbox{\boldmath $\cdot$}}
\newcommand{\bfL}{\mbox{\boldmath $L$}}
\newcommand{\bfp}{\mbox{\boldmath $p$}}
\newcommand{\bfq}{\mbox{\boldmath $q$}}
\newcommand{\bfn}{\mbox{\boldmath $n$}}
\newcommand{\bfP}{\mbox{\boldmath $P$}}
\newcommand{\bfQ}{\mbox{\boldmath $Q$}}
\newcommand{\bfT}{\mbox{\boldmath $T$}}
\newcommand{\bfI}{\mbox{\boldmath $I$}}

\def\aj{{AJ}}                   
\def\araa{{ARA\&A}}             
\def\apj{{ApJ}}                 
\def\apjl{{ApJ}}                
\def\aap{{A\&A}}                
\def\mnras{{MNRAS}}             
\def\prl{{Phys.~Rev.~Lett.}}    
\def\physrep{{Phys.~Rep.}}   

\def\ltsima{$\; \buildrel < \over \sim \;$}
\def\simlt{\lower.5ex\hbox{\ltsima}}
\def\gtsima{$\; \buildrel > \over \sim \;$}
\def\simgt{\lower.5ex\hbox{\gtsima}}
\def\gsim{ \lower .75ex \hbox{$\sim$} \llap{\raise .27ex \hbox{$>$}} }
\def\lsim{ \lower .75ex\hbox{$\sim$} \llap{\raise .27ex \hbox{$<$}} }
\def\gtrsim{ \lower .75ex \hbox{$\sim$} \llap{\raise .27ex \hbox{$>$}} }
\def\lesssim{ \lower .75ex\hbox{$\sim$} \llap{\raise .27ex \hbox{$<$}} }

\def\cm{\hbox{\,cm}}
\def\yr{\hbox{\,yr}}

\def\half{{\frac{1}{2}}}

\def\p{\partial}

\def\bfI{{\bf I}}

\def\kms{\,\mbox{km s}^{-1}}

\def\msun{\,{\rm M}_\odot}
\def\pc{\,{\rm pc}}
\def\arcsec{\,{\rm arcsec}}
\def\kpc{\,{\rm kpc}}
\def\Myr{\,{\rm Myr}}

\def\comment#1{}

\title[Resonant relaxation and warp of the GC disc]{Resonant relaxation and
  the warp of the stellar disc in the Galactic centre}
\author[B. Kocsis \& S. Tremaine]{
Bence Kocsis$^{1,2,3}$\thanks{bkocsis@cfa.harvard.edu}
and Scott Tremaine$^{2}$\thanks{tremaine@ias.edu}\\
$^{1}$ Harvard-Smithsonian Center for Astrophysics,
60 Garden St., Cambridge, MA 02138, USA \\
$^{2}$ Institute for Advanced Study, Princeton, NJ 08540, USA \\
$^{3}$ Einstein Fellow
}

\begin{document}

\date{Received ---}
\maketitle

\begin{abstract}
  Observations of the spatial distribution and kinematics of young stars in
  the Galactic centre can be interpreted as showing that the stars occupy one,
  or possibly two, discs of radii $\sim 0.05$--0.5 pc. The most prominent
  (`clockwise') disc exhibits a strong warp: the normals to the mean orbital
  planes in the inner and outer third of the disc differ by $\sim 60^\circ$.
  Using an analytical model based on Laplace--Lagrange theory, we show that
  such warps arise naturally and inevitably through vector resonant relaxation
  between the disc and the surrounding old stellar cluster.
\end{abstract}

\section{Introduction}

Black holes of mass $10^6$ to $10^9\msun$ are found in the centres of most
galaxies. These exotic objects are the engines that drive quasars and other
active galactic nuclei, and may play an important role in galaxy formation
through feedback to the interstellar medium. Our own Galaxy contains a central
black hole of mass $\sim 4\times10^6\msun$ associated with the radio source
Sgr A*, and thus provides a unique opportunity to explore the interactions of
a nuclear black hole with the surrounding gas and stars, at spatial resolution
far greater than can be achieved for any other galaxy.

Among the more remarkable components of the Galactic nucleus is the population
of $\sim 100$ massive young stars found in the central parsec
($1\pc=26\arcsec$)\footnote{We exclude the separate population of stars
  (`S-stars') found at much smaller radii, $\lesssim 0.5\arcsec\sim 0.02\pc$.}.
These are mostly O supergiants and Wolf-Rayet (WR) stars, with masses $\gtrsim
20\msun$, formed in a burst lasting $\lesssim 2\Myr$ approximately $6\pm2\Myr$
ago \citep{pau06}.  Proper motions and radial velocities are available for
most of these stars \citep{bartko09}.  In the standard description, about half
of the massive stars between $1\arcsec$ and $10\arcsec$ belong to a rotating system (the
`clockwise disc', \citealt{lb03}), which can be modelled as a warped disc
with local thickness (standard deviation of the inclinations) of about
$14^\circ$ and mean eccentricity of 0.3--0.45 \citep{bartko09}. The warp is
substantial: the symmetry axis of the disc varies by $60^\circ$--$70^\circ$
between $1\arcsec$ and $10\arcsec$ radius. About 20\% of the stars appear to be members
of a second rotating system (the `counter-clockwise disc'), which is thicker
than the clockwise disc and inclined by $115^\circ$ to the clockwise disc in
the same radius range. The properties and even the existence of the
counter-clockwise disc are controversial \citep{lu09}, even though its
statistical significance is $7\sigma$ according to \citet{bartko10}.  The
total mass of the two discs is 5--10$\times10^3\msun$ \citep{pau06}. Inside
$1\arcsec=0.04\pc$ there is a sharp cutoff in the density of WR/O stars.  The
surface density of the clockwise disc is
\begin{equation}
\Sigma\propto r^{-\delta}
\label{eq:sd}
\end{equation}
with $\delta\simeq 1.4\pm 0.2$ between $1\arcsec$ and $15\arcsec$ \citep{bartko10}. The
disc(s) are surrounded by a spherical cluster of old stars \citep{buc09}. The
cluster of old stars is much more massive than the disc(s) -- $5\times
10^5\msun$ inside the outer edge of the disc at $10\arcsec$ -- but
still $\lesssim 10\%$ of the mass of the central black
hole, so the disc(s) are nearly Keplerian.

The disc(s) present a number of puzzles:

\begin{itemize}

\item The existence of young, massive stars implies that star formation must
  have occurred recently in the central parsec. This is surprising, since
  tidal forces from the black hole suppress star formation unless the density
  of the protostellar cloud is orders of magnitude larger than currently
  observed in this region \citep{mor93,alex05}.

\item The presence of two distinct discs suggests that there were two separate
  star-formation events. But then why do the stars in the two discs appear to
  have the same age to within $1\Myr$ \citep{pau06}?

\item The complexity of the dynamical models (two intersecting discs, warps,
  etc.) needed to explain the kinematic and spatial distribution of the disc
  stars suggests that some other structure may provide a better description of
  the data. What is the nature of this structure and why is the
  distribution of young stars so complicated?

\end{itemize}

In this paper we shall focus on one aspect of the Galactic-centre disc(s),
their substantial warp. We shall argue that the warp arises from slowly
varying stochastic torques exerted on the disc by the surrounding cluster of
old stars, through the process sometimes called vector resonant relaxation.
Other properties of the disc(s) may also arise through resonant relaxation, a
discussion that we defer to future papers.

There is an extensive literature on warps of stellar and gasous discs
in the galactic context \citep[for general reviews see,
e.g.,][]{binney92,nt96,bt08}.  In subparsec scale accretion discs,
warps can arise through relativistic frame dragging (Lense--Thirring
precession) around a spinning black hole \citep{bp75}, radiation
pressure \citep{petterson77,pringle96}, gravitational torques due to
massive tori such as the molecular torus in the Galactic centre
\citep{nay05,subr09}, a binary companion such as a second black hole
orbiting inside the disc \citep{ptl98,yt03,yu07},
or stochastic torques from a surrounding star cluster \citep{ba09}.
The self-gravity of a stellar disc can play an important role in
determining the shape of the warp \citep{ht69,nt96,uga09}. Using
$N$-body simulations of an isolated, initially flat and thin stellar
disc that resembles the Galactic-centre disc, \citet{cuadra08} showed
that the observed warp cannot arise from interactions among the disc
stars\footnote{This result is consistent with the analytic arguments
  below that two-body and resonant relaxation among the disc stars is
  unimportant.}. \citet{nc05} and \citet{hn09} have suggested that the
Galactic-centre disc(s) could have been formed in a high-inclination
collision between two massive gas clouds.  \citet{lb09} have examined
the interaction of the stellar disc with a possible massive inclined
second stellar disc, and showed that the discs dissolve due to the
mutual torques on timescales comparable to their age.

In this paper, we examine the evolution of an initially thin, flat
disc in a near-Keplerian gravitational potential, accounting for both
the self-gravity of the disc and stochastic gravitational torques from
a surrounding cluster of stars.  We argue that the most important
torques are those that arise from the mass distribution of the cluster
stars after averaging over orbital phase and apsidal angle (vector
resonant relaxation) and that the self-gravity of the disc suppresses
the excitation of small-scale normal modes so that vector resonant
relaxation warps the disc, rather than thickening it.  In
\S\ref{sec:timescale}, we discuss the timescales of various processes
in the Galactic centre, and establish that vector resonant relaxation
with the old cluster stars is significant for the Galactic-centre
disc(s), whereas most other dynamical relaxation processes (e.g.,
scalar resonant relaxation, two-body relaxation, etc.) are not.  In
\S\ref{sec:laplag}, we derive an analytic solution to the time
evolution of an initially flat stellar disc, based on the following
approximations: (i) the orbital period and the apsidal precession
period are much shorter than the timescale for the warping of the disc,
so we can average the motion over the orbital phase and apsidal angle;
(ii) external torques on these timescales are generated exclusively by
the orbit-averaged mass distribution of the cluster stars (i.e., vector resonant relaxation),
(iii) the eccentricities of the disc stars are negligible;
(iv) the warping angle is small (in particular, the relative
inclination between any two disc stars is small compared to their
fractional difference in semi-major axis);
(v) the orbits of cluster stars are uncorrelated and independent of
the disc (i.e., the back-reaction of the disc on the cluster is
negligible); (vi) the cluster is spherical on average (i.e., the
non-spherical component of the cluster mass distribution is due solely
to Poisson fluctuations from individual stars).  In this case, the
problem reduces to Laplace--Lagrange theory, in which the secular
evolution of the disc is described by a quadratic Hamiltonian, and the
disc is equivalent to a system of point masses interconnected
with springs and driven by the external forces from the cluster. This
system is integrable as the
disc can be decomposed into independent normal modes (i.e., independent
harmonic oscillators).  In \S\ref{sec:interactions}, we derive the
stochastic evolution of the normal mode amplitudes and consider
applications to the discs in the Galactic centre. Our conclusions are
discussed in \S\ref{sec:discussion}.  The statistical equilibrium of
an isolated self-gravitating stellar disc is presented in the
Appendix.

In future work we shall present more general numerical models that do
not require the approximation that the inclinations and eccentricities
are small.

\section{Time-scales in the Galactic centre}
\label{sec:timescale}

We now ask which dynamical processes can shape the properties of the disc over
its lifetime.

Two recent estimates of the distance and mass of the black hole in the
Galactic centre are $R_0=8.0\pm0.6\kpc$, $M_\bullet=(4.1\pm0.6)\times
10^6\msun$ \citep{ghez08} and $R_0=8.3\pm0.4\kpc$, $M_\bullet=(4.3\pm0.4)\times
10^6\msun$ \citep{gil09}. For simplicity we shall adopt $R_0=8\kpc$ and
$M_\bullet=4\times10^6\msun$. At this distance $1\pc=25.8\arcsec$ and
$1\arcsec=0.0388\pc$.

\begin{figure*}
\centering
\includegraphics[width=130mm]{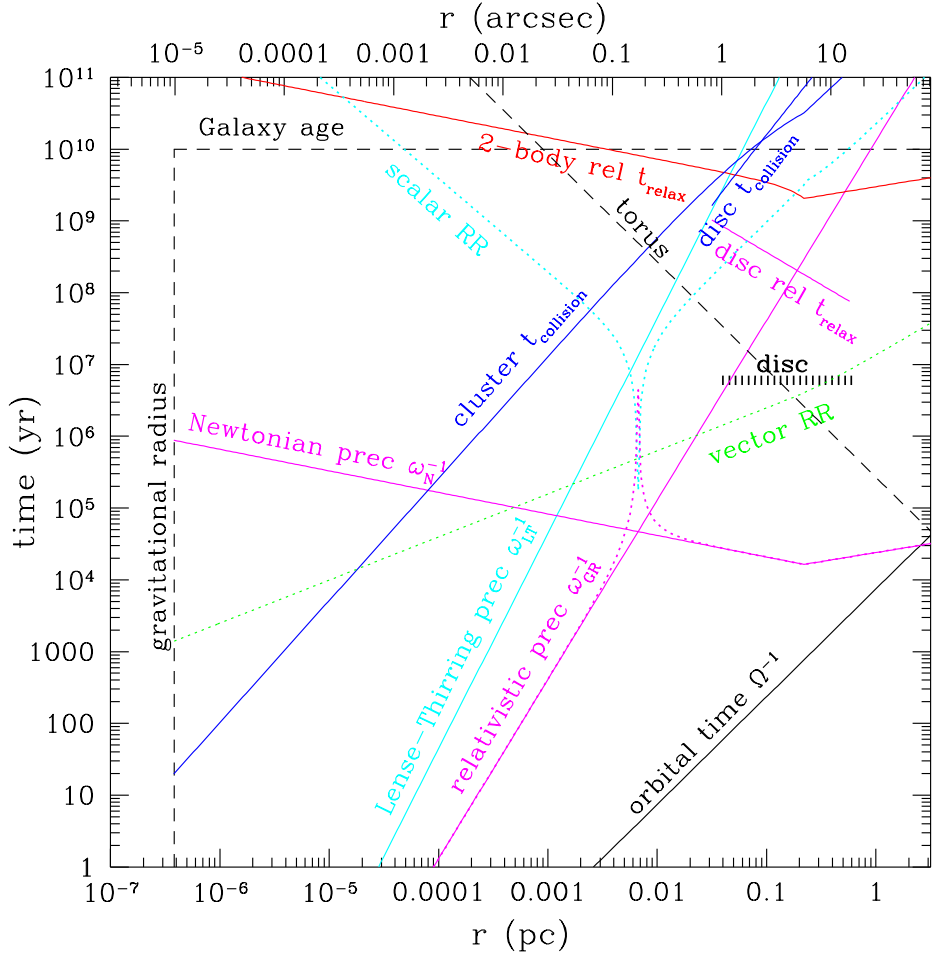}
\caption{Time-scales in the central parsec of the Galaxy. The dashed black
  lines show the gravitational radius of the central black hole (eq.\
  \ref{eq:rgrav}) and the age of the Galaxy. The hatched black line shows the
  age and radial extent of the discs(s) of WR/O stars. The slanted black line
  at the lower right shows the inverse orbital frequency $\Omega^{-1}$ (eq.\
  \ref{eq:omegadef}). The magenta lines show the apsidal precession
  times due to general relativity $\omega_{\rm GR}^{-1}$ (eq.\ \ref{eq:gr})
  and the stellar cluster $|\omega_{N}|^{-1}$ (eq.\ \ref{eq:newt}), and the
  magenta points show the combined apsidal precession time $|\omega_{\rm
    GR}+\omega_{N}|^{-1}$. The solid cyan line slanting up to the right shows
  the Lense--Thirring precession time (eq.\
  \ref{eq:ltdef}). The red line shows the two-body relaxation time for the
  stellar cluster assuming that $m_2=\langle m^2\rangle/\langle
  m\rangle=1\msun$ (eq.\ \ref{eq:trelax}), and a magenta line shows the
  two-body relaxation time within the disc, assuming $m_2=10\msun$ and a disc
  mass of $5000\msun$ (eq.\ \ref{eq:trelaxd}). The dotted lines show the
  scalar and vector resonant relaxation time-scales in cyan and green, for
  $m_2=10\msun$ (eqs.\ \ref{eq:tress} and \ref{eq:tresv}). The solid blue
  lines show the collision time in the stellar cluster for stars with mass
  $m=\msun$ and radius $r_\star=R_\odot$ and in the disc for $m=20\msun$ and
  radius $r_\star=10R_\odot$ (eqs.\ \ref{eq:tcolls} and \ref{eq:tcolld}).
  Finally, the slanted dashed black line shows the precession time due to the
  molecular torus. }
\label{fig:time}
\end{figure*}

The times and distances derived below are plotted in Figure \ref{fig:time},
which also shows the disc age ($6\pm2\Myr$) and radial extent (0.04--$0.6\pc$)
as a hatched bar.

\paragraph*{Gravitational radius} The gravitational radius of the black hole
is\comment{actual number 1.1813}
\begin{equation}
r_\bullet=\frac{2GM_\bullet}{c^2}=1.18\times10^{12}\cm,
\label{eq:rgrav}
\end{equation}
marked in the figure by a dashed vertical line.

\paragraph*{Orbital frequency}
The orbital frequency $\Omega$ of a star with semi-major axis $a$ is given
by\comment{actual number 235.778}
\begin{equation}
\Omega^{-1}=\left(\frac{a^3}{GM_\bullet}\right)^{1/2}=
236\yr\left(\frac{a}{0.1\pc}\right)^{3/2}.
\label{eq:omegadef}
\end{equation}
The characteristic orbital time $\Omega^{-1}$ (orbital period divided by
$2\pi$) is plotted in Figure \ref{fig:time}.

\paragraph*{Relativistic precession}
The apsidal precession rate due to general relativity $\omega_{\rm GR}$ is
given by\comment{actual number 4.10628}
\begin{equation}
\omega_{\rm GR}^{-1}=\frac{c^2a^{5/2}(1-e^2)}{3(GM_\bullet)^{3/2}}=
4.11\times 10^7\yr\,(1-e^2)\left(\frac{a}{0.1\pc}\right)^{5/2}
\label{eq:gr}
\end{equation}
where $e$ is the eccentricity. The characteristic precession time $\omega_{\rm
  GR}^{-1}$ for nearly circular orbits ($e\simeq0$) is plotted in magenta in
Figure \ref{fig:time}.  Over the radius range of the disc(s), the relativistic
precession time (\ref{eq:gr}) is much larger than the Newtonian precession
time (\ref{eq:newt}), also plotted in magenta.  Thus relativistic apsidal
precession is unimportant for the disc(s), although it is likely to dominate
over Newtonian precession inside $0.1\arcsec$.

The orbit-averaged Lense--Thirring precession of a star with angular momentum
$\bfL$ is \citep{ll}
\begin{equation}
\frac{d\bfL}{dt}=\frac{2G^2M_\bullet^2s}{a^3c^3(1-e^2)^{3/2}}\bfn_\bullet\cross\bfL
\equiv \omega_{\rm LT}\,\bfn_\bullet\cross\bfL,
\end{equation}
where $\bfn_\bullet$ is the unit vector parallel to the spin axis of the black
hole and $0\le s<1$ is the spin parameter, that is, the spin angular momentum
of the black hole is $sGM^2/c$. The characteristic precession time
\comment{actual number 4.4522}
\begin{align}
\omega_{\rm LT}^{-1}=&\frac{a^3c^3(1-e^2)^{3/2}}{2G^2M_\bullet^2s}
\notag \\
=&\frac{4.45\times10^{10} \yr}{s}(1-e^2)^{3/2}
  \left(\frac{a}{0.1\pc}\right)^3
\label{eq:ltdef}
\end{align}
is plotted in Figure \ref{fig:time} in cyan, for circular orbits and a
maximally spinning black hole ($s=1$).
Lense--Thirring precession is negligible for the disc(s), at least at the
radii where they are currently observed ($\gtrsim 1\arcsec$).

\paragraph*{The star cluster}

The black hole is surrounded by an approximately spherical cluster of old
stars. The distribution of mass in the cluster can be measured from star
counts by assuming that the mass density is proportional to the number
density. The proportionality constant can be estimated from the kinematics
(radial velocities and proper motions) of stars at radii $\sim1\pc$, where the
stellar mass begins to make a substantial contribution to the overall
gravitational force field.

Using data from \cite{sch07}, \cite{lbk09} estimate that the mass density at
radius $r$ is given by
\begin{equation}
\rho(r)=(2.8\pm 1.3)\times
10^6\msun\pc^{-3}\left(\frac{r}{0.22\pc}\right)^{-\gamma},
\label{eq:rho}
\end{equation}
with $\gamma=1.2$ inside $0.22\pc$ and $\gamma=1.75$ outside $0.22\pc$.
The corresponding enclosed mass is\comment{actual numbers 5.0350,1.1187,0.9158}
\begin{align}
&\frac{M(r)}{\msun}=\frac{4\pi}{\msun}\int_0^r\rho(x)x^2\,dx
\label{eq:mass}  \\ &=
\left\{
\begin{array}{ll}
0.50\times 10^5(r/0.1\pc)^{1.8}, & r < 0.22\pc,
\\
1.12\times 10^5(r/0.1\pc)^{1.25}-0.92\times 10^5,
& r > 0.22\pc,
\end{array}
\right.\notag
\end{align}
with an uncertainty of about 50\%.

For comparison, \cite{tri08} give
\begin{equation}
\rho(r)=2.1\times
10^6\msun\pc^{-3}\frac{1}{1+(r/0.34\pc)^2},
\end{equation}
which yields an enclosed mass
\begin{equation}
\frac{M(r)}{\msun}=1.0\times 10^6\msun \,f\hspace{-2pt}\left(\frac{r}{0.34\pc}\right),
\end{equation}
where $f(x)\equiv x-\tan^{-1}x$.
The two expressions for $M(r)$ agree to within a factor of two for
$r\gtrsim0.3\pc$ but differ by more than an order of magnitude for $r\lesssim
0.05\pc$. At $r=1\pc$ the two expressions give $M(r)=1.9\times10^6\msun$ and
$1.7\times10^6\msun$ respectively, somewhat larger than the independent
estimate of 1.1--$1.5\times10^6\msun$ given by \cite{sch09}. For our numerical
results we shall use the parametrization (\ref{eq:mass}).

\paragraph*{Velocity dispersion}
Assuming that the velocity dis\-per\-sion tensor of the stellar cluster is
approximately isotropic, and solving the Jeans equation for the
one-dimensional velocity dispersion $\sigma(r)$ (eq.\ 4.216 in \citealt{bt08}),
we find\comment{actual numbers 279.62,250.10,0.03529}
\begin{equation}
\sigma(r)=\left\{\begin{array}{ll}
280\kms\sqrt{0.1\pc/r}\,\sqrt{1-0.035\left(r/0.1\pc\right)^{2.2}}, \\
&\hspace{-80pt} {\rm if~} r < 0.22\pc, \\[1.25ex]
250\kms\sqrt{0.1\pc/r}, &\hspace{-80pt} {\rm if~} r > 0.22\pc.
\end{array}
\right.
\label{eq:sigma}
\end{equation}

\paragraph*{Newtonian precession} The apsidal precession rate $\omega_{\rm N}$
due to the gravitational field from a spherical star cluster is always
negative, that is, the line of apsides precesses in the opposite direction to
the orbital motion \citep[e.g.,][]{tre05}. For a cluster with density $\rho(r)$ and
mass $M(r)\ll M_\bullet$, the precession rate of a nearly circular orbit with
radius $r$ is $\omega_N=-2\pi G\rho(r)/\Omega(r)$, or\footnote{This formula was
  derived for eccentricity $e\ll1$, but works fairly well for $e\lesssim
  1$. For example, if the cluster density is a power law in radius, $\rho(r)\propto
  r^{-\gamma}$ with $1<\gamma<3$, the error is less than 30\% for all
  eccentricities $e<0.7$.}\comment{actual numbers 2.0812,1.3489}
\begin{equation}
|\omega_{\rm N}|^{-1}=\left\{\begin{array}{ll}
2.1\times 10^4\yr\,(r/0.1\pc)^{-0.3}, &{\rm if~} r < 0.22\pc \\
1.3\times 10^4\yr\,(r/0.1\pc)^{0.25}, &{\rm if~} r > 0.22\pc, \end{array}
\right.
\label{eq:newt}
\end{equation}
plotted as a magenta line in Figure \ref{fig:time}. We also plot as magenta
dots the total precession rate $\omega\equiv \omega_{\rm GR}+\omega_{\rm N}$.
Over the radial extent of the disc(s), the characteristic precession time
$|\omega|^{-1}$ is dominated by Newtonian effects and equal to a few times
$10^4\yr$, more than two orders of magnitude shorter than the disc age.

\paragraph*{Two-body (non-resonant) relaxation}

The two-body relaxation time for the cluster of old stars is given by equation
(7.106) in \cite{bt08},\comment{actual numbers 3.633,1.685}
\begin{equation}
t_{\rm relax}=0.34\,\frac{\sigma^3}{G^2\rho m_2\ln\Lambda}\notag\
\end{equation}
\begin{equation}
\;=\left\{\begin{array}{ll}
3.6\times 10^9\yr\,\left(r/0.1\pc\right)^{-0.3}\left(15/\ln\Lambda\right)(\msun/m_2)&\\
\quad\quad\times\left[1-0.035\left(r/0.1\pc\right)^{2.2}\right]^{3/2},
&\hspace{-50pt}  {\rm if~} r < 0.22\pc \\[1.5ex]
1.7\times 10^9\yr\,\left(r/0.1\pc\right)^{0.25}(15/\ln\Lambda)(\msun/m_2), &\\
   &\hspace{-50pt}{\rm if~} r > 0.22\pc,
\end{array}
\right.
\label{eq:trelax}
\end{equation}
where $\ln\Lambda\simeq \ln(M_\bullet/m)\simeq 15$, and the effective mass
$m_2\equiv \langle m^2\rangle/\langle m\rangle$ is the ratio of the
mean-square stellar mass to the mean stellar mass; we shall call this the
effective mass. The relaxation time $t_{\rm relax}$ for $\ln\Lambda=15$ and
$m_2=\msun$ is plotted in red in Figure \ref{fig:time}.

Unfortunately the effective mass is quite uncertain. We assume a
stellar mass function of the form
\begin{equation}
dn\propto m^{-\alpha}dm\quad \mbox{for $m_{\rm
  min}<m<m_{\rm max}$;}
\label{eq:imf}
\end{equation}
then if $m_{\rm min}\ll m_{\rm max}$,
\begin{equation}
m_2=\left\{\begin{array}{ll}
(\alpha-2)(\alpha-3)^{-1}m_{\rm min} & \alpha >3; \\
(\alpha-2)(3-\alpha)^{-1}m_{\rm max}^{3-\alpha}m_{\rm min}^{\alpha-2} & 2
  < \alpha < 3; \\
(2-\alpha)(3-\alpha)^{-1}m_{\rm max} & \alpha <2.\end{array} \right.
\label{eq:mf}
\end{equation}
Thus for a standard Salpeter mass function ($\alpha=2.35$) with $m_{\rm
  max}=100\msun$ and $m_{\rm min}=0.1\msun$, $m_2=4.8\msun$. For the
\cite{ktg93} solar neighborhood mass function, $m_2=0.66\msun$. However, there
is evidence that the mass function in the Galactic centre is much more
top-heavy than in the solar neighborhood. \cite{bartko10} suggest that the
Galactic-centre disc(s) have $\alpha\simeq0.45$, in which case $m_2=0.6m_{\rm
  max}\sim 60\msun$. Even if the initial mass function were known, there are
other complications. First, the initial-final mass function -- the relation
between the main-sequence stellar mass and the mass of the compact object that
remains after stellar evolution is complete -- is poorly known.
Second, star clusters, massive gas clouds, and other density inhomogeneities also
contribute to $m_2$ and may lead to values of $m_2$ that are orders of magnitude
larger than the typical stellar mass \citep{perets07}.
An example in the Galactic Centre is IRS 13E, a compact cluster of three
bright blue supergiants and many fainter components with a velocity dispersion
suggesting a central mass of a few $10^4 \msun$ \citep{fritz10}.

A further uncertainty is that two-body relaxation may lead to mass segregation
so that the effective mass depends on radius. In particular, the relaxation
time inside a few tenths of a parsec may be dominated by stellar remnants
rather than stars \citep{okl09}. If the mass function is broad and dominated by light
stars, then the heavy stars develop a much steeper cusp with $\rho\propto r^{-2}$
to $r^{-3}$. This affects $m_2$ primarily very close to the black hole, increasing
$m_2$ by a factor of $\sim 2$ at 0.4\,pc and a factor $\sim 4$ at 0.04\,pc \citep{ah09,kha09}.

The two-body relaxation time for $m_2=1\msun$ shown in Figure \ref{fig:time}
is never shorter than a few Gyr so even if $m_2\sim 10^2\msun$ two-body
relaxation is negligible over the age of the disc stars, $\lesssim
10\Myr$. However, two-body relaxation may have a substantial effect on old
stars in the surrounding cluster.

For massive stars such as the WR/O stars observed in the disc(s),
dynamical friction from the star cluster can act on a shorter time-scale than
$t_{\rm relax}$.  The time-scale for orbital decay of a star of mass $m$ on a
near-Keplerian circular orbit of radius $r$ is given by
\begin{equation}
t_{\rm fric}^{-1}\equiv \frac{1}{r}\frac{dr}{dt}=8\pi\Omega \frac{m}{ M_\bullet}
\frac{\rho r^3}{ M_\bullet} \ln\Lambda\, g(X),
\label{eq:fricdef}
\end{equation}
where $g(X)=\mbox{erf}(X)-X\mbox{erf}\,'(X)$ and $X=v/(\sqrt{2}\sigma)$ where
$v$ is the circular speed at $r$ \citep{bt08}\footnote{This result assumes
  that the distribution function is Maxwellian but should be approximately
  valid for more realistic distribution functions.}. If the density varies as a
power law with radius, $\rho\propto r^{-\gamma}$, then
$X=[\half(1+\gamma)]^{1/2}$. Using $v=(GM_\bullet/r)^{1/2}$ and the dispersion
profile (\ref{eq:sigma}), we find that $g(X)$ varies from 0.47 for
$r\ll0.2\pc$ to 0.57 for $r\gg0.2\pc$ so for our purposes it is adequate to
use a constant value $g(X)=0.5$. With this approximation\comment{actual
  numbers 1.734,1.124}
\begin{equation}
t_{\rm fric}=\left\{\begin{array}{ll}
1.7\times 10^8\yr\,\left(r/0.1\pc\right)^{-0.3}(12/\ln\Lambda)(20\msun/m), & \\
 & \hspace{-70pt}\hbox{if~}r < 0.22\pc,
\\[2ex]
1.1\times 10^8\yr\,\left(r/0.1\pc\right)^{0.25}(12/\ln\Lambda)(20\msun/m), & \\
 & \hspace{-70pt}\hbox{if~} r > 0.22\pc. \end{array}\right.
\label{eq:friction}
\end{equation}
We conclude that the action of dynamical friction from the star cluster on the
disc stars is negligible over the disc age of $6\Myr$, even for stars as
massive as $100\msun$.

The effects of two-body relaxation between the disc stars are uncertain
because the properties of the disc(s) are poorly determined and the relaxation
time is a strong function of the root mean squared (rms) eccentricity and
inclination. For our purposes it is sufficient to approximate the disc as a
single population of stars of mass $m$, and in this case the eccentricity
relaxation time is given by \citep{si00}
\begin{equation}
t_{\rm relax}^{-1}\equiv \frac{1}{ \langle e^2\rangle}
\frac{d \langle e^2\rangle}{dt}
=4.5\frac{\Omega}{\langle e^2\rangle^2}
\frac{m}{M_\bullet}
\frac{\Sigma r^2}{M_\bullet}\ln\Lambda.
\end{equation}
Here $\langle e^2\rangle$ is the mean-square eccentricity of the disc stars,
$\Sigma(r)$ is the surface density, and $\Lambda\simeq \langle
e^2\rangle^{3/2}M_\bullet/m$. This formula assumes that the rms
inclination $\langle i^2\rangle^{1/2}\simeq 0.5\langle e^2\rangle^{1/2}$, a
typical value seen in relaxed discs and consistent with observations of the
clockwise disc; and that $\Lambda\gg1$ (the disc is `dispersion-dominated'). The
inclination relaxes at a rate
\begin{equation}
\frac{1}{\langle i^2\rangle}
\frac{d \langle i^2\rangle}{dt}
=\frac{0.5}{t_{\rm relax}}.
\end{equation}
We assume that the surface-density distribution in the disc is given by
equation (\ref{eq:sd}) and parametrize the disc by its total mass $M_{\rm
  disc}$, which is probably about $5000\msun$
\citep{pau06}. Then\comment{actual number 3.65}
\begin{equation}
t_{\rm relax} =
4\times 10^8\yr
\frac{\langle e^2\rangle^2}{(0.3)^4}
\frac{5000\msun}{M_{\rm disc}}
\frac{10\msun}{m_2}
\left(\frac{0.1\pc}{r}\right)^{0.9}\!\!
\frac{9}{\ln\Lambda},
\label{eq:trelaxd}
\end{equation}
where $\ln\Lambda=9.3$ corresponds to $m=10\msun$, $\langle
e^2\rangle^{1/2}=0.3$. Given these parameters, the disc relaxation time is
plotted in magenta in Figure \ref{fig:time}. Even for a large effective mass,
$m_2\sim 10^2\msun$, the minimum relaxation time over the radial range of the
disc(s) is $\gtrsim 3\times 10^7\yr$, so two-body relaxation should be
negligible for most stars over the disc age of $6\Myr$. However, if the
eccentricities and inclinations in the disc have been over-estimated, perhaps
because of unrecognized systematic errors, the scaling $t_{\rm relax}\sim e^4$
implies that the relaxation time could be much shorter.

\paragraph*{Physical collisions} For a population of stars of a single mass,
the rate of physical collisions is \citep[][eq.\ 7.195]{bt08}
\begin{equation}
t_{\rm coll}^{-1} = 16\sqrt{\pi} n\sigma r_\star^2(1+\Theta)
\label{eq:tcolls}
\end{equation}
where $n=\rho/m$ is the number density of stars and $\Theta\equiv
\frac{1}{4}v_\star^2/\sigma^2$ where $v_\star=\sqrt{2Gm/r_\star}$ is the
escape speed from the surface of the star, radius $r_\star$. The collision
time of stars in the cluster is shown in blue in Figure \ref{fig:time},
assuming that the cluster is mainly composed of solar-type stars with
$m=\msun$ and $v_\star=618\kms$. At radii $\lesssim 0.01\pc$, where the
collision time is less than a few Gyr, collisions are
likely to play a dominant role in determining the distribution of old
solar-type stars and the giants into which they eventually evolve, which have
much shorter lifetimes but much larger radii. Collisional destruction of
red-giant envelopes may be responsible for the depletion in luminous red
giants observed inside $\sim 1\pc$ and the disappearance of the CO spectral
feature associated with red giants in the integrated light
\citep{alex05,dale09}.  However, collisions with stars in the old cluster do not
play a major role in the evolution of the early-type stars in the disc(s),
since the collision time is much longer than the disc age,
even after accounting for the larger radii and masses of the WR/O stars.

In a dispersion-dominated disc composed of identical stars of mass $m$ and
radius $r_\star$, the collision rate is \citep[e.g.,][]{ht09}
\begin{equation}
t_{\rm coll}^{-1} = 16\frac{\Sigma}{m}\Omega r_\star^2(0.69+1.52\Theta), \quad
  \Theta=\frac{m}{M_\bullet}\frac{r}{r_\star}\frac{1}{\langle e^2\rangle}.
\end{equation}
With the same parameters used to derive equation (\ref{eq:trelaxd}), we find
\comment{actual number 4.05}
\begin{equation}
t_{\rm coll} = \frac{4.1\times10^{11}\yr}{1+2.20\Theta}\,
\frac{5000\msun}{M_{\rm disc}}
\frac{m}{20\msun}
\left(\frac{r}{0.1\pc}\right)^{2.9}\!
\left(\frac{10R_\odot}{r_\star}\right)^2 \notag
\end{equation}
\begin{equation}
\Theta = 24.6\frac{m}{20\msun}
\frac{10R_\odot}{r_\star}
\frac{r}{0.1\pc}
\frac{(0.3)^2}{\langle e^2\rangle}.
\label{eq:tcolld}
\end{equation}
This result is shown as a blue line in Figure \ref{fig:time}. The collision
time is longer than the disc age.

\paragraph*{Scalar resonant relaxation} Resonant relaxation \citep{rt96} arises
from the forces due to the orbit-averaged mass distribution of the stars.
Because these forces vary only slowly, they affect the angular momentum but
not the energy of stellar orbits. Scalar resonant relaxation affects the
magnitude of the angular momentum and thus the eccentricity, while vector
resonant relaxation affects the direction of the angular momentum vector but
not its magnitude. The scalar resonant relaxation time in a spherical stellar
system is \citep{ha06}
\begin{equation}
t_{\rm rr,s}=
\frac{4\pi|\omega|}{\beta_s^2\Omega^2}
\frac{M_\bullet^2}{M(r)m_2};
\label{eq:tress}
\end{equation}
here $\omega$ is the apsidal precession rate, the sum of the (negative)
Newtonian rate $\omega_{\rm N}$ (eq.\ \ref{eq:newt}) and the (positive)
relativistic rate $\omega_{\rm GR}$ (eq.\ \ref{eq:gr}). The dimensionless
coefficient $\beta_s$ is estimated to be $1.05\pm0.02$ by \cite{eil09}. Using
these parameters and effective mass $m_2=10\msun$ we plot the scalar resonant
relaxation time using cyan dots in Figure \ref{fig:time}. The cuspy minimum
near $r=0.007\pc$ arises where relativistic precession cancels
Newtonian precession so the total precession rate vanishes\footnote{The
    sharpness of this minimum is artificial, since
the total precession rate vanishes at different semi-major axes for orbits of
different eccentricities \citep{gh07}.}.

The scalar resonant relaxation time is less than $10^{10}\yr$ throughout the
central parsec. Thus the eccentricity distribution of old stars in this region
should be relaxed. However, the scalar resonant relaxation time exceeds
$10^8\yr$ throughout the radial extent of the disc(s), so eccentricity relaxation
of the disc stars is likely to be small.
However, it is likely that the eccentricity relaxes faster for the most
massive stars in the cluster, similar to how dynamical friction
(\ref{eq:fricdef}) is faster than two-body relaxation for massive stars \citep{rt96}.
Assuming that the analogous resonant dynamical friction time-scale is
inversely proportional to stellar mass\footnote{The rate of 
resonant dynamical friction can be computed
using the formalism for ordinary dynamical friction derived by
\citet{tw84}, in which the friction arises from stars that are in
near-resonance with an orbiting massive body; resonant friction is
stronger than ordinary friction because some of the resonant
denominators are near zero.}, it is still comparable to the disc age
only near its inner edge.

Scalar resonant relaxation can also occur among the stars in the disc(s). The
rate for this process is more difficult to determine \citep{tre98}.

\paragraph*{Vector resonant relaxation}
This process affects the orientation of the angular-momentum vector but not
its magnitude. In a stellar system in which the time-averaged gravitational
field is spherically symmetric, the vector resonant relaxation time is
\citep{eil09}\footnote{We have set Eilon et al.'s parameter $A_\phi$ to
  unity.}
\begin{equation}
t_{\rm rr,v}=
\frac{2\pi}{\beta_v^2\Omega}M_\bullet
\frac{1}{\sqrt{M(r) m_2}}.\label{eq:tresvdef}
\end{equation}
Adopting $\beta_v=1.83\pm0.03$ from \cite{eil09}, we have\comment{numbers
  7.8857,5.2905}
\begin{equation}
t_{\rm rr,v}=\left\{\begin{array}{ll}
7.9\times 10^6\yr\,\sqrt{\msun/ m_2}(r/0.1\pc)^{0.6}, & \\
&\hspace{-20pt}{\rm if~} r < 0.22\pc, \\
5.3\times 10^6\yr\,\sqrt{\msun/ m_2}(r/0.1\pc)^{0.9}\\
\ \times\left[1-0.82(r/0.1\pc)^{-1.25}\right]^{-1/2}\!\!\!,
&\hspace{-20pt}{\rm if~} r > 0.22\pc.\end{array}
\right.
\label{eq:tresv}
\end{equation}
This result is plotted in green in Figure \ref{fig:time}, for $m_2=10\msun$.
At all radii, vector resonant relaxation is substantially faster than scalar
resonant relaxation.  The inclination distribution of old stars should be
relaxed throughout the stellar cluster.

Equation (\ref{eq:tresv}) is only valid if the nodal precession rate is
dominated by the stochastic component of the non-spherical gravitational force.
Hence this equation may overestimate the relaxation rate if the stellar system
is not spherically symmetric. For example, the non-spherical field from the disc
is smaller than the stochastic field from the cluster stars only if $M_{\rm
  disc} \lesssim [M(r)m_2]^{1/2}$. For $M(r)\sim 1\times 10^5\msun$ (cf.\ eq.\
\ref{eq:mass}) and $m_2\sim 100\msun$ (see discussion following eq.\
\ref{eq:mf}), this requires $M_{\rm disc}\lesssim 3\times 10^3\msun$, compared
to an estimated mass of $5\times 10^3\msun$. Thus the approximation that the
precession rate is dominated by stochastic forces is suspect, and should be
improved in future models.

Vector resonant relaxation may be responsible for warps in gaseous accretion
discs in the centres of galaxies, in particular for the warp of the maser disc
in the galaxy NGC 4258 \citep{ba09}.  It may also be the principal mechanism that
isotropizes the inclinations of S-stars in the Galactic centre
\citep{ha06,perets09}. In this paper, we investigate whether vector resonant
relaxation leads to the warped structure of the stellar disc in the
Galactic centre \citep{bartko09}.

Vector resonant relaxation can also occur among the stars in the disc. Nodal
precession within the disc is determined by the mean gravitational field of
the flattened mass distribution, rather than the random torques from
individual stars. The typical nodal precession rate for a star in the disc
is\footnote{Note that the rate diverges as the thickness of the disc approaches zero,
because the torque exerted on a ring by a massive, razor-thin disc approaches
infinity near the disc.
}
\begin{equation}
\nu \simeq
\frac{\Omega}{\langle i^2\rangle^{1/2}}
\frac{M_{\rm disc}}{M_\bullet}.
\end{equation}
We then estimate the vector resonant relaxation time from equation
(\ref{eq:tress}) by replacing the apsidal precession rate $\omega$ with the
nodal precession rate $\nu$ and the cluster mass $M(r)$ with the disc mass
$M_{\rm disc}$:
\begin{align}
t_{\rm rr,v}\simeq& \frac{4\pi}{\Omega\langle i^2\rangle^{1/2}}
\frac{M_\bullet}{m_2}\notag\\
\simeq& 2.4\times 10^9\yr
\frac{14^\circ}{\langle i^2\rangle^{1/2}}
\frac{20\msun}{m_2}
\left(\frac{r}{0.1\pc}\right)^{3/2}.
\label{eq:tresvd}
\end{align}
This is much longer than both the age of the disc(s) and the vector
resonant relaxation time-scale due to the cluster stars. Thus vector resonant
relaxation among the disc stars is unlikely to play an important role in
determining the properties of the Galactic-centre
disc(s)\footnote{Nevertheless, it is worthwhile to understand the effects of
  vector resonant relaxation on an old, isolated disc, and this is the subject
  of the Appendix.}.

\paragraph*{Warping by a massive perturber}
The disc(s) can be warped by a massive perturber, such as the molecular torus
outside the disc at 1.5--7 pc \citep{chris05} or an intermediate-mass black
hole inside the disc \citep{yu07}. This process was first investigated by
Laplace in the context of planetary satellites and has been studied by many
authors since then \citep{ht69,nt96,uga09}; see \cite{nay05}, \cite{lb09}, and
\cite{subr09} in the context of the Galactic-centre disc(s). The precession
rate of a circular ring at radius $r$ due to a circular ring of mass $m_p$ at
radius $r_p$ with relative inclination $I$ is given by equation
(\ref{eq:torque}); in the limit where $r\ll r_p$ or $r\gg r_p$ the result
simplifies to \citep{nay05}
\begin{equation}
 \nu_p \simeq -\frac{3}{4}\Omega \frac{m_p}{M_{\bullet}} \frac{r
   r_<^2}{r_>^3}\cos I,
\label{eq:nayak}
\end{equation}
where $r_<=\min(r,r_p)$, $r_>=\max(r,r_p)$. This formula
assumes that the mass of the perturbed ring is sufficiently small that its
angular momentum is much less than the angular momentum of the perturber.

The characteristic precession time $t_p=1/\nu_p$ for an external perturber
($r\ll r_p$) is then\comment{exact number 8.496}
\begin{equation}
 t_p \simeq 8.5 \times 10^6 \yr \frac{10^6\msun}{M_p}
 \left(\frac{r_p}{1.5\pc}\right)^{3} \left(\frac{0.1\pc}{r}\right)^{3/2}
\frac{0.5}{\cos I}.
\end{equation}
The reference masses and radii have been set to resemble the molecular torus
\citep{chris05}. For these parameters, the precession time-scale is longer than
the age of the disc at $r\lesssim 0.1\pc$.  However, for $r\gtrsim 0.1\pc$,
the molecular torus can warp the Galactic-centre disc(s) significantly, at
least if $\cos I$ is not too small (Fig.\ \ref{fig:time}). Curiously, the
molecular torus {\em is} nearly orthogonal to the mean orientation of the
clockwise stellar disc ($\cos I \simeq 0.00 \pm 0.03$). More precisely it is
nearly orthogonal to the inner parts ($\cos I \simeq 0.05 \pm 0.03$ for
$r\lesssim 0.15\pc$), but not the outer parts ($\cos I\simeq 0.52$ and 0.24
between $r=0.13$--$0.27\pc$ and 0.27--$0.46\pc$, respectively).
\citet{subr09} argued that the molecular torus could lead to the thickening
and warping of the stellar disc if it were initially not orthogonal.

The precession time for an internal perturber ($r\gg r_p$) is\comment{exact
  number 5.696}
\begin{equation}
 t_p \simeq 5.7\times 10^6 \yr \frac{5000\msun}{M_p}
 \left(\frac{0.1\pc}{r_p}\right)^{2} \left(\frac{r}{0.2\pc}\right)^{7/2}
\frac{0.5}{\cos I}.
\end{equation}
The reference masses and radii have been set to resemble the counter-clockwise
disc \citep{bartko09}. For these parameters the precession time-scale is
comparable to the age of the disc; in fact, \cite{ndcg06} have used
$N$-body simulations to set an upper limit of $5\times10^3\msun$ on the mass
of the counter-clockwise disc from the requirement that it does not warp the
clockwise disc too much.

\bigskip We have seen that a rich set of internal and external
dynamical processes can affect the evolution and current state of the Galactic
centre disc(s). A useful first approximation is to neglect external tidal
fields, two-body relaxation, dynamical friction, scalar resonant relaxation,
and scalar resonant friction with the surrounding old stellar cluster, as well
as two-body and resonant relaxation between the disc stars. We
shall also assume that there is no massive perturber in the disc. We may then  focus on
the effects of vector resonant relaxation with the surrounding cluster.  In
the next sections we develop analytic machinery to provide a description of
this interaction.

\section{Laplace--Lagrange theory for an isolated disc}

\label{sec:laplag}

Vector resonant relaxation affects the orientation of stellar orbits but
not their semi-major axes or eccentricities. Since the relaxation time is much
larger than the apsidal precession time (\ref{eq:newt}) each orbit may be
thought of as an axisymmetric planar annulus obtained from the Keplerian orbit
by averaging over mean anomaly and argument of
pericentre.

We investigate the dynamics of the disc using a simple model system. To
construct this we shall assume that the stellar orbits in the disc are nearly
coplanar ($I\ll 1$) and nearly circular ($e\ll1$). In fact we shall make an
even stronger assumption: that $e,I\ll \Delta a/a$ where $\Delta a$ is the
typical difference in semi-major axis between a star and its nearest neighbor.
This condition is satisfied in many planetary systems, including our own, but
probably not in the Galactic-centre disc(s); nevertheless we
shall argue below that it allows an analytic treatment of resonant relaxation
that captures its most important features.

With these approximations, we evaluate the Hamiltonian of a system of $N$
stars of masses, semi-major axes, inclinations, and nodes $m_i$, $a_i$, $I_i$,
and $\Omega_i$, $i=0,\ldots,N-1$, to quadratic order in the eccentricity and
inclination. This is the classical Laplace--Lagrange secular theory
\citep{md99}. In this theory the inclination and node are decoupled from the
eccentricity and apse. The evolution of the latter two elements is scalar
resonant relaxation, which we have argued is unimportant because of the
relatively rapid apsidal precession induced by the stellar cusp. The evolution
of the former is vector resonant relaxation; the relevant Hamiltonian is
\begin{align}
H(\bfq,\bfp)=&\frac{G}{8}\sum_{i=0}^{N-1}\sum_{j=i+1}^{N-1}
\frac{m_im_j}{\max(a_i,a_j)}
\alpha_{ij}b_{3/2}^{(1)}(\alpha_{ij})\notag\\
&\quad\times\left[\left(\frac{p_i}{\gamma_i}-\frac{p_j}{\gamma_j}\right)^2
+\left(\frac{q_i}{\gamma_i}-\frac{q_j}{\gamma_j}\right)^2\right],
\label{eq:hamsec}
\end{align}
where $\alpha_{ij}=\min(a_i,a_j)/\max(a_i,a_j)$,
\begin{equation}
b_{3/2}^{(1)}(\alpha)=\frac{2}{\pi}\int_0^\pi
\frac{\cos\theta\,d\theta}{(1-2\alpha\cos\theta+\alpha^2)^{3/2}}
\end{equation}
is the Laplace coefficient\footnote{In terms of the complete elliptic
integrals of the first and second kind
\begin{equation}
K(k) = \int_0^{\pi} \frac{d\theta}{\sqrt{1-k^2 \sin^2 \theta}} ,\quad E(k) = \int_0^{\pi} \sqrt{1-k^2 \sin^2 \theta}  \,d\theta,
\end{equation}
the Laplace coefficient satisfies
\begin{equation}
 b_{3/2}^{(1)}(\alpha) = \frac{4}{\pi \alpha (1-\alpha^2)^2}
\left[   (1+\alpha^2)E(\alpha) - (1-\alpha^2)K(\alpha)  \right].
\end{equation}
}, $\gamma_i=m_i^{1/2}(GM_\bullet a_i)^{1/4}$, and
\begin{equation}
q_i\equiv \gamma_iI_i\sin\Omega_i,\quad
p_i\equiv -\gamma_iI_i\cos\Omega_i
\label{eq:pqdef}
\end{equation}
are canonical coordinates and momenta. For some purposes it is useful to
`soften' the Laplace coefficient to suppress the singularity at $\alpha=1$,
replacing the formula above by
\begin{equation}
b_{3/2}^{(1)}(\alpha,\epsilon)=\frac{2}{\pi}\int_0^\pi
\frac{\cos\theta\,d\theta}{(1-2\alpha\cos\theta+\alpha^2+\epsilon^2)^{3/2}}
\label{eq:soft}
\end{equation}
where $\epsilon$ is the dimensionless softening parameter and  $0<\epsilon\ll1$.

We use $(x,y,z)$ to denote the standard Cartesian coordinates relative to
which the node $\Omega$ and inclination $I$ are measured.
The total angular momentum is
\begin{align}
\left(\begin{array}{c}
L_x\\
L_y\\
L_z
\end{array}
\right)
=&\sum_{i=0}^{N-1} m_i\sqrt{GM_{\bullet} a_i}
\left(\begin{array}{c}
\sin I_i\sin\Omega_i\\
-\sin I_i\cos\Omega_i\\
\cos I_i
\end{array}
\right)
 \notag \\
=&\sum_{i=0}^{N-1} m_i\sqrt{GM_{\bullet} a_i}
\left(\begin{array}{c}
I_i\sin\Omega_i\\
-I_i\cos\Omega_i\\
1-\half I_i^2
\end{array}\right)
+\hbox{O}(I_i^3) \notag \\
=&\sum_{i=0}^{N-1}
\left(\begin{array}{c}
\gamma_iq_i\\
\gamma_i p_i\\
\gamma_i^2-\half q_i^2-\half p_i^2
\end{array}\right)
+\hbox{O}(I_i^3) .
\label{eq:jdef}
\end{align}
It is straightforward to show that all three components of the total angular
momentum are conserved (see discussion following eqs.\
\ref{eq:xxx}--\ref{eq:zdeff}).

The Hamiltonian can be rewritten as
\begin{equation}
H(\bfq,\bfp)=\bfp^{\rm T}\bfssA\bfp + \bfq^{\rm T}\bfssA\bfq,
\end{equation}
where $\bfp^{\rm T}=(p_0,\ldots,p_{N-1})$, $\bfq^{\rm T}=(q_0,\ldots,q_{N-1})$ and the
$N\times N$ matrix $\bfssA$ is defined by
\begin{align}
A_{ij}=&
 -\frac{Gm_im_j\alpha_{ij}}{8\,\max(a_i,a_j)\gamma_i\gamma_j}
b_{3/2}^{(1)}(\alpha_{ij}) \qquad \mbox{ if $i\not=j$} \notag \\
 =&\sum_{k\not=i}
\frac{Gm_im_k\alpha_{ik}}{8\,\max(a_i,a_k)\gamma_i^2}
b_{3/2}^{(1)}(\alpha_{ik}) \qquad \mbox{if $i=j$.}
\label{eq:adef}
\end{align}
Note that
\begin{equation}
\sum_{j=0}^{N-1} A_{ij}\gamma_j=0.
\label{eq:constraint}
\end{equation}

Since $\bfssA$ is symmetric, it can be diagonalized in the form
\begin{equation}
\bfssA=\bfssO\bfssLambda\bfssO^{\rm T},
\label{eq:ortho}
\end{equation}
where $\bfssO$ is orthogonal ($\bfssO^{\rm T}=\bfssO^{-1}$) and $\bfssLambda$ is
diagonal.  The diagonal elements of $\bfssLambda$ are the eigenvalues of
$\bfssA$ and the columns of $\bfssO$ are the normalized eigenvectors (see
Figures \ref{fig:Lambdahist}--\ref{fig:Modes} below for properties of the
eigenvalues and eigenvectors in simulated discs).

Now consider a canonical transformation to new coordinates and momenta
$(\bfQ,\bfP)$ having the generating function
\begin{equation}
S=\bfP^{\rm T}\bfssO^{\rm T}\bfq.
\end{equation}
Then
\begin{equation}
\bfp=\frac{\p S}{\p\bfq}=\bfssO\bfP,\quad  \bfQ=
\frac{\p S}{\p\bfP}=\bfssO^{\rm
  T}\bfq,\quad  \bfq=\bfssO\bfQ,
\label{eq:matrix}
\end{equation}
and
\begin{equation}
H(\bfP,\bfQ)=\bfP^{\rm T}\bfssLambda\bfP+\bfQ^{\rm T}\bfssLambda\bfQ.
\label{eq:orthoham}
\end{equation}
The inclinations of stars can be calculated from $\bfQ$ and $\bfP$ as
\begin{equation}
I_j^2= \frac{q_j^2 + p_j^2}{\gamma_j^2} = \left(\sum_{i=0}^{N-1} \frac{O_{ji}}{\gamma_j} Q_i \right)^2 +
\left(\sum_{i=0}^{N-1} \frac{O_{ji}}{\gamma_j} P_i \right)^2.
\label{eq:I}
\end{equation}
Since $H$ is separable, $P_i^2+Q_i^2$ is a constant of motion for all
$i=0,\ldots,N-1$. The equations of motion are
\begin{equation}
  \dot Q_i=
\frac{\p H}{\p P_i}=2\Lambda_iP_i, \quad \dot P_i=
-\frac{\p H}{\p Q_i}=-2\Lambda_iQ_i,
\label{eq:eqmot}
\end{equation}
which have the solution
\begin{align}
  Q_i(t)=& Q_i(0)\cos(2\Lambda_it)+P_i(0)\sin(2\Lambda_it),\notag\\
  P_i(t)=&-Q_i(0)\sin(2\Lambda_it)+P_i(0)\cos(2\Lambda_it).
\end{align}
Thus $\Lambda_i$ can be regarded as a frequency.

Since $b_{3/2}^{(1)}(\alpha)$ is positive, equation (\ref{eq:hamsec}) implies
that $H\ge0$. Thus $H(\bfq,\bfp)$ is a positive semi-definite
quadratic form defined by the matrix $\bfssA$, and thus all of the eigenvalues
$\Lambda_i$ are non-negative. The lower bound $H=0$ is achieved if
$p_i/\gamma_i=\hbox{const}$, $q_i/\gamma_i=\hbox{const}$ independent of $i$,
corresponding to a
uniform tilt of the disc plane. Thus there is one zero eigenvalue, which we
call $\Lambda_0$, and the rest are positive; in the discussions below we
always order the eigenvectors by eigenvalue, so $\Lambda_0=0 < \Lambda_1 <
\cdots < \Lambda_{N-1}$. The eigenvector corresponding to
zero eigenvalue is $c(\gamma_0,\ldots,\gamma_{N-1})$ where $c$ is a constant
(cf.\ eq.\ \ref{eq:constraint}). Since the columns of $\bfssO$ are normalized
eigenvectors, $O_{j0}=c\gamma_j$ with $c=\pm(\sum_j\gamma_j^2)^{-1/2}$ and
\begin{align}
Q_0=&\sum_{j=0}^{N-1}O_{j0}q_j=c\sum_{j=0}^{N-1}\gamma_jq_j=cL_x, \notag\\
P_0=&\sum_{j=0}^{N-1}O_{j0}p_j=c\sum_{j=0}^{N-1}\gamma_jp_j=cL_y,
\label{eq:xxx}
\end{align}
where the last equalities follow from (\ref{eq:jdef}). Since $\Lambda_0=0$,
$P_0$ and $Q_0$ are constant, so the $x$ and $y$ components of the total
angular momentum are conserved, as they must be.

We define
\begin{equation}
Z(\bfq,\bfp)\equiv \sum_{i=0}^{N-1} \big(q_i^2+p_i^2\big).
\label{eq:zdef}
\end{equation}
Following \cite{las00}, we call $\half Z$ the `angular-momen\-tum
deficit' since it represents the difference between the $z$ component of the
angular momentum of the actual system and the angular momentum that it would
have if all of the stars were on coplanar circular orbits (eq.\
\ref{eq:jdef}). Since the canonical transformation from $(\bfq,\bfp)$ to
$(\bfQ,\bfP)$ is orthogonal
\begin{equation}
Z(\bfQ,\bfP)=\sum_{i=0}^{N-1} \big(Q_i^2+P_i^2\big)=\bfQ^{\rm T}\bfQ+\bfP^{\rm
  T}\bfP.
\label{eq:zdeff}
\end{equation}
Since $Q_i^2+P_i^2$ is constant for all $i$, $Z$ is also constant, which
confirms that the $z$ component of angular momentum is conserved, as it must
be.

Normal modes that have wavelengths that are short compared to the disc radius
but long compared to the radial separation between stars (roughly, $a \gg
\lambda \gg a/N$ where $a$ is the disc radius and $N$ is the number of stars)
approximately satisfy the WKB dispersion relation for bending waves \citep{ht69,bt08}
\begin{equation}
(\omega-m\Omega)^2=\nu^2+ (2\pi)^2\frac{G\Sigma}{\lambda}
\end{equation}
where $\nu$ is the vertical frequency arising from an external potential and
$m$ is the azimuthal wavenumber. In our case $m=1$, $\nu=\Omega$, and
$|\omega|\ll\Omega$, so the dispersion relation simplifies to
\begin{equation}
\omega=-{2\pi^2G\Sigma\over\Omega\lambda}.
\end{equation}
The $n^{\rm th}$ normal mode has approximately $n$
nodes, and so the average wavelength of this mode is roughly $2 (r_{\max}-r_{\min})/n$.
The longest waves have the lowest frequencies, and the frequency of mode $n$
is proportional to $\omega_n\propto n$, consistent with the low-frequency behavior seen in
Figure \ref{fig:Lambdahist} below.

\section{Interactions between the disc and a surrounding stellar system}
\label{sec:interactions}

Let us suppose that there is an external perturbation to the disc, which can
be represented by generalized forces $f_{qi}$, $f_{pi}$ that change the coordinate
and momentum of ring $i$ at a rate $\dot q_i=f_{qi}$, $\dot p_i=f_{pi}$.
Since $\gamma_iq_i$ and $\gamma_ip_i$ are the $x$ and $y$ components of the
angular momentum of star $i$ (cf.\ eq.\ \ref{eq:jdef}) we have
\begin{equation}
f_{qi}(t)=\gamma_i^{-1}T_{xi}(t), \quad f_{pi}(t)=\gamma_i^{-1}T_{yi}(t)
\label{eq:f}
\end{equation}
where $T_{xi}(t)$, $T_{yi}(t)$ are the $x$ and $y$ components of the
orbit-averaged external torque on star $i$. If these torques arise from a
surrounding distribution of stars that is stationary and spherical on average,
then $\langle f_{qi}(t)\rangle=\langle f_{pi}(t)\rangle=0$ for all $i$ where
$\langle\cdot\rangle$ denotes an ensemble average over realizations of the
surrounding stellar cluster.
With the same assumptions
we also have that $\langle f_{qi}(t)f_{pj}(t')\rangle=0$ for all $i$ and $j$.
Since the number of
stars in the surrounding cluster is large, $f_{qi}(t)$ and $f_{pi}(t)$ are
Gaussian random processes. These can be uniquely characterized by the correlation coefficient
\begin{equation}
\Gamma_{ij}(t,t') \equiv \Gamma_{ij}(|t - t'|) \equiv \langle f_{qi}(t)f_{qj}(t')\rangle = \langle f_{pi}(t)f_{pj}(t') \rangle,
\label{eq:correlation}
\end{equation}
where the first and last equality are valid if the forces are
stationary and spherically symmetric on average. Note that
$\Gamma_{ij}(t)$ need not be diagonal since nearby stars are expected
to experience similar torques. The coherence time $\tau$ of the
external torques is defined so that
\begin{equation}
\Gamma_{ij}(\Delta t) \simeq 0 {\rm~~for~~} |\Delta t| \gsim \tau.\label{eq:tau}
\end{equation}
In the next two subsections, we discuss the general properties of the growth of
perturbations from an initially thin coplanar disc, then we construct a specific model for
the torques and correlations.

\subsection{Response of the disc to external torques}\label{sec:growth}
The equations of motion (\ref{eq:eqmot}) in the canonical coordinates
$(\bfQ,\bfP)$ are modified to
\begin{equation}
  \dot Q_i=2\Lambda_iP_i+ \sum_{j=0}^{N-1}O_{ji}f_{qj}(t),\
  \dot  P_i=-2\Lambda_iQ_i+ \sum_{j=0}^{N-1}O_{ji}f_{pj}(t).
\label{eq:eqmota}
\end{equation}
Let us assume that the modes of the disc are non-degenerate,
that is, $\Lambda_i=\Lambda_j$ only if $i=j$ (this is not an important
restriction for practical purposes). With the initial conditions
$Q_i(0)=P_i(0)=0$, equations (\ref{eq:eqmota}) have the solution
\begin{align}
  Q_i(t)=&\sum_{j=0}^{N-1}O_{ji} \int_0^t dt_1\big\{f_{qj}(t_1)
\cos[2\Lambda_i(t-t_1)]\notag \\
&\qquad\qquad+f_{pj}(t_1)\sin[2\Lambda_i(t-t_1)]\big\}
\notag \\
  P_i(t)=&\sum_{j=0}^{N-1}O_{ji} \int_0^t dt_1\big\{f_{pj}(t_1)
\cos[2\Lambda_i(t-t_1)]\notag \\
&\qquad\qquad -f_{qj}(t_1)\sin[2\Lambda_i(t-t_1)]\big\}.
\label{eq:eqmotb}
\end{align}
The mean squared value of the process over different realizations of the
perturbing torques is
\begin{align}
&\langle Q_i^2(t)\rangle = \langle P_i^2(t)\rangle \notag\\
&= \sum_{n,m=0}^{N-1}O_{ni}O_{mi} \int_0^t dt_1\int_0^t dt_2 \notag\\
&\quad \times\big\{\langle f_{qn}(t_1)f_{qm}(t_2)\rangle
\cos[2\Lambda_i(t-t_1)]\cos[2\Lambda_i(t-t_2)]\notag\\
&\quad+\langle f_{pn}(t_1)f_{pm}(t_2)\rangle \sin[2\Lambda_i(t-t_1)]\sin[2\Lambda_i(t-t_2)]\big\}\notag\\
&= \sum_{n,m=0}^{N-1}O_{ni}O_{mi}  \int\limits _0^t dt_1\int\limits_0^t dt_2 \, \Gamma_{nm}(t_2-t_1)\cos[2\Lambda_i(t_2-t_1)]\notag\\
&= 2\sum_{n,m=0}^{N-1}O_{ni}O_{mi}\int_0^t dt'\; (t-t')\Gamma_{nm}(t')\cos 2\Lambda_i t'
\label{eq:QQ}
\end{align}
where in the second-last line we have used the definition of $\Gamma_{nm}(t)$
(eq.~\ref{eq:correlation}) and the trigonometric identity for the sum of
cosines, and in the last line we have used $\Gamma_{nm}(t)=\Gamma_{nm}(-t)$.
Note also that
\begin{equation}
\langle Q_i(t)P_i(t)\rangle=0.
\label{eq:QQPP}
\end{equation}

Similarly, we can calculate the cross-correlation coefficient between two modes
with $\Lambda_i \neq \Lambda_j$,
\begin{align}
&\langle Q_i(t) Q_j(t)\rangle =\langle P_i(t) P_j(t)\rangle  \notag\\
&=\sum_{n,m=0}^{N-1}O_{ni}O_{mj} \int_0^t dt_1\int_0^t dt_2 \notag   \\
  &\quad \times \big\{ \langle f_{qn}(t_1)f_{qm}(t_2)\rangle\cos[2\Lambda_i(t-t_1)]
\cos[2\Lambda_j(t-t_2)]\notag\\
&\quad+\langle f_{pn}(t_1)f_{pm}(t_2)\rangle\sin[2\Lambda_i(t-t_1)]\sin[2\Lambda_j(t-t_2)]\big\}
\notag\\
&= \sum_{n,m=0}^{N-1}O_{ni}O_{mj}\int_0^t dt_1\int_0^t dt_2   \notag\\
&\quad\times\Gamma_{nm}(t_2-t_1)\cos[2\Lambda_i(t-t_1)
-2\Lambda_j (t-t_2)]\notag\\
&= \sum_{n,m=0}^{N-1}O_{ni}O_{mj}
\frac{\cos(\Delta_{ij}t)}{\Delta_{ij}}\int_0^t dt'\;\Gamma_{nm}(t')\notag\\
&\quad\times\big[
\sin(\Delta_{ij}t-2\Lambda_i t') + \sin(\Delta_{ij}t + 2\Lambda_j t')
\big]
\label{eq:QQQQQ}
\end{align}
where we have introduced $\Delta_{ij}=\Lambda_i - \Lambda_j$
to simplify notation.

A similar calculation shows that
\begin{align}
\langle Q_i(t) P_j(t)\rangle = \tan(\Delta_{ij}t)\langle Q_i(t) Q_j(t)\rangle.
\label{eq:cross}
\end{align}

The mean squared inclinations at time $t$ follow from equations (\ref{eq:I}),
(\ref{eq:QQ}), and (\ref{eq:QQQQQ}),
\begin{align}
\langle I_j^2(t) \rangle  =& \sum_{k,l=0}^{N-1} \frac{O_{jk} O_{jl}}{\gamma_{j}^2}
\left[ \langle Q_k(t) Q_l(t)\rangle + \langle P_k(t) P_l(t)\rangle\right]
\label{eq:IRMS}
\end{align}

In the rest of this paper we explore the implications of equations
(\ref{eq:eqmota})--(\ref{eq:IRMS})  for the evolution of an initially
planar disc excited by stochastic torques. First, we make general
statements, then we examine applications to the Galactic-centre disc(s) in the
next section.

\begin{figure}
\centering
\includegraphics[width=84mm]{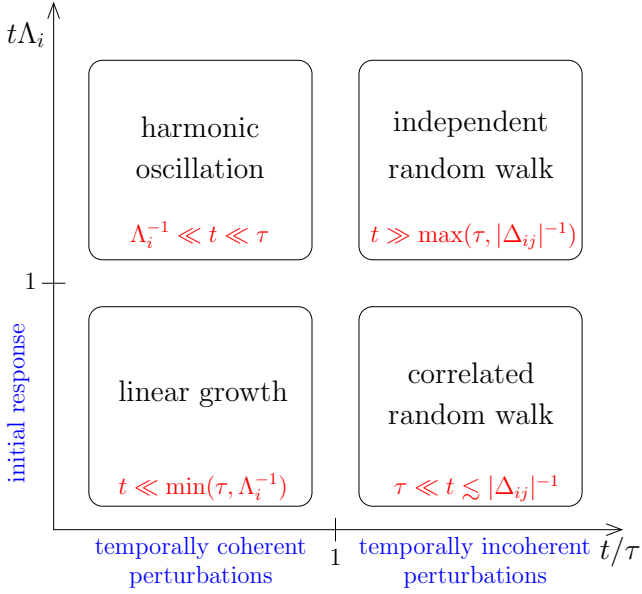}
\caption{Evolutionary stages of the excitation of normal modes of an
  initially thin, flat stellar disc by stochastic external
  perturbations.  On the horizontal axis, the age of the disc $t$ is
  measured relative to $\tau$, the coherence time of the perturbations.
 The vertical axis shows $t$ relative to
  the inverse frequencies $\Lambda_i^{-1}$ of the normal modes of the
  disc.  If $t\ll \tau$ the perturbations are temporally coherent, and
  the normal mode amplitudes initially grow linearly, then saturate and
  oscillate with angular frequency $2\Lambda_i$.  For $t \gg \tau$
  the perturbations are temporally incoherent, and the
  amplitudes of modes $i$ and $j$ undergo initially correlated, later
  independent, random walks (before and after time
  $|\Delta_{ij}^{-1}|=|\Lambda_i - \Lambda_j|^{-1}$, respectively).
  The evolution in these regimes is examined
  separately in \S\ref{sec:initialresponse}, \ref{sec:coherent}, and
  \ref{sec:incoh}.  }
\label{fig:regimes}
\end{figure}
These equations contain three characteristic time-scales: (i) the age
of the disc $t$, which determines the interval over which the external
torques have acted on the initially thin and flat disc; (ii) the inverse
frequencies $\Lambda_i^{-1}$ of the normal modes of the disc;
(iii) the coherence time $\tau$ of the perturbations (eq.\ \ref{eq:tau}).

Note that any perturbation that is initially zero, and then grows and decays
smoothly with coherence time $\tau$ (i.e., a perturbation whose temporal power
spectrum contains only frequencies $\gtrsim 1/\tau$) cannot excite short-period normal modes
($\Lambda_i^{-1}\ll \tau$) because the actions and hence amplitudes
of these modes are adiabatic invariants (an analogy arises in
the disruption of open clusters by molecular clouds: the tidal forces from the
clouds are ineffective if the orbital period of the stars in the cluster is
small compared to the passage time; see \citealt{spi58}).

The time evolution of the disc can be
understood analytically in the following limiting cases,
shown schematically in Figure \ref{fig:regimes}.

\subsubsection{Initial response, $t\ll \Lambda_i^{-1}$}
\label{sec:initialresponse}

When the age of the disc is much less than the inverse frequency of a
particular normal mode $i$, the factor $\cos 2\Lambda_i t'$ in the
integrand of equation (\ref{eq:QQ}) is unity and we have
\begin{equation}
\langle Q_i^2\rangle =\langle P_i^2\rangle=
2\sum_{n,m=0}^{N-1}O_{ni}O_{mi}\int_0^t dt'\; (t-t')\Gamma_{nm}(t').
\label{eq:initialgrowth-Q}
\end{equation}
In particular, the zero-frequency ($i=0$) normal mode describing the
mean orientation of the disc (cf.\ eq.\ \ref{eq:xxx}) always grows
according to equation (\ref{eq:initialgrowth-Q}).

If the disc age is much smaller than $\Lambda_i^{-1}$ for {\em all}
modes $i$, then equation (\ref{eq:QQQQQ}) yields
\begin{equation}
\langle Q_iQ_j\rangle =\langle P_iP_j\rangle=
2\sum_{n,m=0}^{N-1}O_{ni}O_{mj}\int_0^t dt'\; (t-t')\Gamma_{nm}(t').
\end{equation}
and the orthogonal transformation to $(q_i, p_i)$ yields
\begin{equation}
\langle q_i^2 \rangle = \langle p_i^2 \rangle =
2 \int_0^t dt' (t-t')\Gamma_{ii}(t'),\quad{\rm if}~t\ll \min_i \Lambda_i^{-1}.
\label{eq:initialgrowth-q}
\end{equation}
(The same result would obtain if the disc mass were small enough that the
collective effects from its self-gravity were negligible.)
The inclination distribution is given by equation (\ref{eq:I}),
\begin{equation}
\langle I_j^2 \rangle = \frac{4}{\gamma_j^{4}} \int_0^t dt' (t-t')
\langle T_{x  j}(0)T_{x  j}(t')\rangle,\; t\ll \min_i\Lambda_i^{-1}.
\label{eq:initialgrowth-I}
\end{equation}

\subsubsection{Temporally coherent perturbations, $t\ll \tau$}
\label{sec:coherent}

When the disc age $t$ is much shorter than the coherence time $\tau$,
the external forces are approximately constant over the lifetime of
the disc, so $\Gamma_{jk}(t)\simeq\hbox{const}$, and
equations (\ref{eq:QQ}) and (\ref{eq:QQQQQ}) simplify to
\begin{align}
\langle Q_i^2\rangle=&\langle P_i^2\rangle =\!
 \sum_{n,m=0}^{N-1}O_{ni}O_{mi}\Gamma_{nm}\frac{\sin^2\Lambda_i
   t}{\Lambda_i^2},
\notag\\
\langle Q_iQ_j\rangle=&\langle P_iP_j\rangle =\!
\sum_{n,m=0}^{N-1}O_{ni}O_{mj}\Gamma_{nm}
\frac{\sin\Lambda_i   t}{\Lambda_i}\frac{\sin\Lambda_j t}{\Lambda_j} \cos\Delta_{ij}t.
\label{eq:correlatedgrowth}
\end{align}
Initially, the amplitude of each normal mode grows linearly with time
with a rate independent of $\Lambda_i$.
Then after a saturation time $t_{{\rm s},i}\equiv \half\pi/\Lambda_i$ the amplitude
reaches a maximum and begins to oscillate. At much larger times the
time-averaged mean-square amplitude is $\langle Q^2_i\rangle = \half \sum_{nm} O_{ni} O_{mi}
\Gamma_{nm}/ \Lambda_i^2$.

The time evolution for any particular realization of the perturbing forces
can be obtained directly from equation (\ref{eq:eqmotb}) by
specializing to constant $f_{qj}$ and $f_{pj}$,
\begin{align}
  Q_i(t)=&\sum_{j=0}^{N-1}\frac{O_{ji}}{2\Lambda_i}\big[f_{qj}\sin2\Lambda_it
-f_{pj}(\cos2\Lambda_it-1)\big],
\notag \\
  P_i(t)=&\sum_{j=0}^{N-1}\frac{O_{ji}}{2\Lambda_i}\big[f_{pj}\sin2\Lambda_it
+f_{qj}(\cos2\Lambda_it-1)\big]
\label{eq:eqmotc}
\end{align}
for $i>0$ ($\Lambda_i\not=0$); otherwise for $i=0$ ($\Lambda_0=0$)
\begin{equation}
 Q_0(t)=t\sum_{j=0}^{N-1}O_{j0}f_{qj}, \quad
 P_0(t)=t\sum_{j=0}^{N-1}O_{j0}f_{pj}.
\label{eq:eqmotz}
\end{equation}

\subsubsection{Temporally incoherent perturbations, $t\gg\tau$.}

\label{sec:incoh}

When the disc age is much larger than the coherence time,
then since $\Gamma_{nm}(t')=0$ for $t'\gsim\tau$,
the upper limit of the integration in equations (\ref{eq:QQ}) and (\ref{eq:QQQQQ})
can be truncated at $\tau$, so
\begin{align}
\langle Q^2_i(t)\rangle =& \langle P^2_i(t)\rangle = g_it -h_i,\quad{\rm if}~t\gsim\tau,
\label{eq:uncorrelatedgrowth}\\
\langle Q_i(t)Q_j(t)\rangle =& \frac{\cos \Delta_{ij}t}{\Delta_{ij}}\, \left[A_{ij} \cos {\Delta_{ij}}t + B_{ij} \sin {\Delta_{ij}}t\right],\notag\\
\langle Q_i(t)P_j(t)\rangle =& \frac{\sin \Delta_{ij}t}{\Delta_{ij}}\, \left[A_{ij} \cos {\Delta_{ij}}t + B_{ij} \sin {\Delta_{ij}}t\right],\notag
\end{align}
where the constants $A_{ij}$, $B_{ij}$, $g_i$, and $h_i$ are
\begin{align}
A_{ij} =& \sum_{n,m=0}^{N-1}O_{ni}O_{mj}\int_0^{\tau}\!\! dt'\; \Gamma_{nm}(t')
\left[ \sin(2\Lambda_j t') - \sin (2\Lambda_i t') \right] \notag \\
B_{ij} =& \sum_{n,m=0}^{N-1}O_{ni}O_{mj}\int_0^{\tau}\!\! dt'\; \Gamma_{nm}(t')
\left[ \cos(2\Lambda_i t') + \cos (2\Lambda_j t') \right] \notag \\
g_i =& B_{ii},\quad
h_i = 2\sum_{n,m=0}^{N-1}O_{ni}O_{mi}\int_0^{\tau} dt'\; t'
\Gamma_{nm}(t')\cos 2\Lambda_i t'.
\end{align}

For $t\gg \max(\tau, h_i/g_i)$, $\langle Q^2_i(t)\rangle=
\langle P^2_i(t)\rangle\simeq g_it$; moreover $\langle Q_i(t)P_i(t)\rangle=0$
(eq.\ \ref{eq:QQPP}), and $\langle Q_i(t)Q_j(t)\rangle$ and
$\langle Q_i(t)P_j(t)\rangle$ are bounded by $\pm\sqrt{A_{ij}^2 + B_{ij}^2}/|\Delta_{ij}|$,
so after $t \gg \max\left[\tau, |\Delta_{ij}|^{-1} (A_{ij}^2 + B_{ij}^2)^{1/2}/g_i\right]$,
the cross-correlation becomes negligible compared to $\langle Q^2_i(t)\rangle$.
In other words, at large times $Q_i(t)$ and $P_i(t)$ undergo
random walks.
The random walks of various modes $i$ and $j$ are first correlated then
gradually become independent after the disc lifetime becomes much larger
than the inverse relative normal mode frequencies
$|\Delta_{ij}|^{-1}$.

\bigskip In our case none of these simple limits applies: the coherence time
of the perturbations from the cluster is the vector resonant relaxation
time-scale, and this can be shorter or longer than the disc age depending on
the radius and the effective mass $m_2$ in the cluster (cf.\ Fig.\
\ref{fig:time}).  Moreover the set of inverse frequencies $\Lambda_i^{-1}$
includes values that are both shorter and longer than the disc age (Fig.\
\ref{fig:Lambdahist}).  We compare these time-scales numerically for Monte
Carlo simulations of the stellar disc in \S\ref{sec:monte} below.

Finally, the configuration of an isolated disc in thermal equilibrium
under vector resonant relaxation is discussed in the Appendix.

\subsection{Distribution of torques}

\label{sec:torquedist}

We now construct a model for the perturbing torques
$\{T_{xi},T_{yi}\}$. These are the sum of the torques from all of the
stars in the cluster on the disc star labelled by $i$, in a circular orbit near the $z=0$
plane; the torques are averaged over the orbit of both the cluster
star and the disc star. To simplify the calculation at modest cost in
realism, we assume that the cluster stars are also on circular
orbits. Then the total torque on disc stars on circular orbits can be
calculated from the gravitational torque acting between circular rings
or wires of uniform density:
\begin{equation}
\bfT_i=\sum_\beta \sum_{\ell=1}^\infty  K_{i\beta \ell}
P'_{2\ell}\big(\bfn_{\beta} \bcdot \bfn_{i}\big) \bfn_{i} \cross \bfn_\beta,
\label{eq:torque}
\end{equation}
where $i$ and $\beta$ label disc and cluster stars, respectively,
${\bfL}_i$ or $\bfL_\beta$ is the angular momentum vector of star $i$
or $\beta$, and $\bfn_\beta=\bfL_\beta/|\bfL_\beta|$ is the unit vector aligned
with the angular momentum, which is related to the node
$\Omega_\beta$ and inclination $I_\beta$ by
$\bfn_\beta=(\sin I_\beta\sin\Omega_\beta,-\sin I_\beta\cos\Omega_\beta,\cos
I_\beta)$. We also have
\begin{equation}
 K_{i\beta \ell} =
G m_i m_\beta [P_{2\ell}(0)]^2
\frac{r_{i\beta<}^{2\ell}}{r_{i\beta>}^{2\ell+1}}, \quad
P_{2\ell}(0)=\frac{(-1)^\ell\Gamma(2\ell+1)}{2^{2\ell}\Gamma^2(\ell+1)}.
\end{equation}
Here $m_\beta$ and $r_\beta$ are the mass and orbital radius of star $\beta$
(recall that both the disc and cluster stars are assumed to be on circular
orbits), $r_{i\beta<}=\min(r_i,r_\beta)$, $r_{i\beta>}=\max(r_i,r_\beta)$, and $P_{2\ell}(x)$
and $P'_{2\ell}(x)$ are the Legendre polynomial and its first
derivative\footnote{In terms of the associated Legendre function $P_n^m(x)$,
  $P'_{2\ell}(\cos I) = -P^1_{2\ell}(\cos I)/\sin
  I$.}.  The sum over $\ell$ in equation (\ref{eq:torque}) is absolutely
convergent unless $r_i=r_\beta$, The analytical formula (\ref{eq:nayak}) for
the precession rate when $r_</r_> \ll 1$ is the corresponding
limiting case of equation (\ref{eq:torque}).

In the discussions below, we neglect the back-reaction of the disc stars on
the orbits of stars in the old cluster. This approximation should be valid so
long as the total angular momentum contained in the fluctuating non-spherical
component of the star cluster exceeds the angular momentum in the disc. The
former is roughly $N^{1/2}m_2(GM_\bullet r)^{1/2}$ where $m_2=\langle
m^2\rangle/\langle m\rangle$ is the effective mass and $N=M(r)/m_2$ is the
effective number of stars in the cluster at radius $r$; the latter is $M_{\rm
  disc}(GM_\bullet r)^{1/2}$. Thus the neglect of the back-reaction should be
valid so long as $[M(r)m_2]^{1/2}\gtrsim M_{\rm disc}$. This is the same as
the criterion that nodal precession is dominated by the stochastic field from
the old cluster rather than the disc, as discussed following equation
(\ref{eq:tresv}).

Assuming that the disc is flat (i.e., zero inclination for all disc stars),
equation (\ref{eq:torque}) simplifies to
\begin{equation}
\left(\begin{array}{l}
T_{xi}\\
T_{yi}
\end{array}\right)
=\sum_\beta \sum_{\ell=1}^\infty  K_{i\beta \ell} P^1_{2\ell}(\cos I_{\beta})
\left(\begin{array}{l}
\cos \Omega_{\beta}\\
\sin \Omega_{\beta}
\end{array}\right).
\label{eq:T0}
\end{equation}

\subsubsection{The torques as a Gaussian random process}

\label{sec:correlation}

In a spherical cluster, or the equatorial plane of an axisymmetric cluster,
the torques described by equation (\ref{eq:T0}) are the sum of a large number
of independent random variables with zero mean. Hence, according to the central limit theorem, the
probability distribution of the torques $T_{xi}$ and $T_{yi}$ is Gaussian,
with zero mean and covariance or correlation function $\big\langle T_{xi}(0)
T_{xj}(t)\big\rangle = \big\langle T_{yi}(0)
T_{yj}(t)\big\rangle=\gamma_i\gamma_j\Gamma_{ij}(t)$.  From equation
(\ref{eq:T0}) it is also clear that $\big\langle T_{xi} T_{yj}\big\rangle =0$
in a spherical cluster, for all $i$ and $j$.

A Gaussian random process is fully characterised by its mean and
correlation function.  Thus, when generating a Monte Carlo distribution of
cluster stars, the distribution of torques is the same for all mass
functions having the same effective mass $m_2$ so long as the total
mass distribution $M(r)$ is fixed. Thus we may assume without loss of
generality that all the cluster stars have mass $m_{\beta}=m_2$.

Next we examine the spatial correlation function of cluster torques
$\big\langle T_{xi}(0)
T_{xj}(0)\big\rangle=\gamma_i\gamma_j\Gamma_{ij}(0)$. We are mostly interested
here in the correlation function within a thin, flat disc. Then we may use
equation (\ref{eq:T0}) for the torques, which yields
\begin{align}
\langle T_{xi}(0)&T_{xj}(0)\rangle   \notag \\
=& \left\langle\sum_{\beta,\ell,\ell'} K_{i\beta \ell}K_{j\beta \ell'}
P^1_{2\ell}(\cos I_{\beta}) P^1_{2\ell'}(\cos I_{\beta}) \sin^2 \Omega_{\beta}\right\rangle
\notag\\
=&  G^2 m_i m_j
\sum_{\beta}m_{\beta}^2\sum_{\ell=1}^\infty  [P_{2\ell}(0)]^2 k_{\ell}
\frac{r_{i\beta<}^{2\ell}\,r_{j\beta<}^{2\ell}}{r_{i\beta>}^{2\ell+1}\,r_{j\beta>}^{2\ell+1}},
\label{eq:sigmaij0}
\end{align}
where $k_{\ell}$ is a constant defined as
\begin{align}
 k_{\ell} =&  [P_{2\ell}(0)]^2 \langle P^1_{2\ell}(\cos I_{\beta})
 P^1_{2\ell}(\cos I_{\beta})\rangle
\langle\sin^2 \Omega_{\beta}\rangle  \notag\\
=& \frac{2\ell(1+ 2\ell)}{\pi(1+ 4\ell)}\frac{\Gamma^2(\ell+1/2)}{\Gamma^2(\ell+1)}.
\end{align}
Here we have used the orthogonality of the Legendre functions,
$\langle P^m_{2\ell}(\cos I_{\beta}) P^m_{2\ell'}(\cos I_{\beta})\rangle = 0$ for $\ell\neq \ell'$.

We have separated $k_\ell$ from the factor $[P_{2\ell}(0)]^2$ in equation
(\ref{eq:sigmaij0}) because $k_{\ell}$ is practically independent of $\ell$;
as $\ell$ varies from 1 to $\infty$, $k_\ell$ varies only from $3/10=0.3$ to
$1/\pi=0.3183$.  Therefore we can estimate the correlation function to reasonable
accuracy by replacing $k_\ell$ with a constant $k$, $0.3\le k <0.32$. This
leads to a power series in $\ell$ which is reminiscent of the power series
defining the complete elliptic integral of the first kind, $K(x)
=\half\pi\sum_{n=0}^{\infty} [P_{2n}(0)]^2 x^{2 n}$.  Therefore
\begin{align}
\langle T_{xi}(0)T_{xj}(0)\rangle &  \notag \\
& \hspace{-40pt}= G^2 k m_i m_j \sum_{\beta}
\frac{m_{\beta}^2}{r_{i\beta>}r_{j\beta>}} \!\left[ \frac{2}{\pi}K(\alpha_{i\beta} \alpha_{j\beta})-1\right],
\label{eq:sigmaij}
\end{align}
where $\alpha_{i\beta} = r_{i\beta<} /
r_{i\beta>}=\min(r_i,r_\beta)/\max(r_i,r_\beta)$.  We may
express this result in terms of the
density of the cluster $\rho(r)$ (eq.\ \ref{eq:rho}) and the effective mass
$m_2= \langle m_{\beta}^2\rangle/\langle m_{\beta} \rangle$,
\begin{align}
\langle T_{xi}(0)T_{xj}(0)\rangle =&    4\pi G^2 k m_i m_j m_2 \notag\\
& \hspace{-40pt}\times\int\limits_0^{\infty}\frac{dr\,
  r^2\rho(r)}{\max(r,r_i)\max(r,r_j)}
\left[\frac{2}{\pi}K(\alpha_{i}\alpha_{j})-1\right],
\label{eq:sigmaij2}
\end{align}
where $\alpha_i=\min(r,r_i)/\max(r,r_i)$.

Equations (\ref{eq:sigmaij})--(\ref{eq:sigmaij2}) define the probability
density of torques acting on the disc stars at each instant.  They
show that the correlation function for disc stars at radii $r_i$ and
$r_j$ is determined mainly by cluster stars in the region
$\min(r_i,r_j)\lesssim r \lesssim \max(r_i,r_j)$. In
particular, at large radii $\rho(r)\propto r^{-1.75}$ (eq.~\ref{eq:rho}) so
for $r\gg \max(r_i,r_j)$ the integrand declines as $r_i^2 r_j^2 r^{-5.75}$, so the
contribution to the correlation function from this region is negligible.  At small radii,
$\rho(r)\propto r^{-1.2}$ so for $r\ll \min(r_i,r_j)$ the integrand scales as
$r^{4.2}/(r_i^3 r_j^3)$, implying that the contribution from small radii is also
negligible. \citet{gh07} reach a similar conclusion.

A simple fitting formula for equation (\ref{eq:sigmaij2}), using the
mass distribution in the Galactic centre (eq.\ \ref{eq:mass}), is
\begin{align}
\langle T_{xi}(0)T_{xj}(0)\rangle \simeq&
\,c_T  \frac{m_2 \rho(\bar{r}_{ij})\bar{r}_{ij}^{3} }{ M_{\bullet}^2}L_i L_j
\Omega(r_i)\Omega(r_j){\alpha}_{ij}^{\kappa}
\label{eq:sigmaij3}
\end{align}
where $\alpha_{ij}=\min(r_i,r_j)/\max(r_i,r_j)$,
$\bar{r}_{ij}=\sqrt{r_i r_j}$, $L_i=m_i r_i^2\Omega(r_i)$ is the
angular momentum, and $c_T$ and $\kappa$ are fit parameters.  The best
fit values, which approximate equation (\ref{eq:sigmaij2}) to better
than $20\%$ over radii $\alpha_{ij}<4$ and $\bar{r}_{ij}<4\pc$, are
$c_T=0.77$ and $\kappa=1.8$.  Thus, the correlation of torques on
different stars, $\langle T_{xi}T_{xj}\rangle/[\langle
T_{xi}^2\rangle^{1/2}\langle T_{xj}^2\rangle^{1/2}]$, is less than
$14\%$ for $a_i \geq 3a_j$. Furthermore, the distribution of torques is a smooth function
of radii with a peak at $a_i=a_j$.

\section{Monte Carlo simulations}
\label{sec:monte}

We construct a flat, razor-thin disc of $N$ stars on circular orbits. The disc
is assumed to lie initially in the reference plane so
$I_i=q_i/\gamma_i=p_i/\gamma_i=0$, for all $i=0,\ldots,N-1$. The semi-major
axes are chosen randomly from the surface density distribution (\ref{eq:sd})
with exponent $\delta=1.4$ \citep{bartko10}, between inner and outer radii of
0.04 and $0.6\pc$ (roughly $1\arcsec$ to $15.5\arcsec$). The stellar masses are chosen
from the mass function (\ref{eq:imf}) with $\alpha=-0.45$, $m_{\rm
  max}=30\msun$ and $m_{\rm min}=1\msun$; the minimum mass is arbitrary but
this choice has almost no influence on our results, and the maximum is the
most massive star that can survive for the 6 Myr age of the disc(s)
\citep{lj01}.  We set the total number of stars to be 500 in this mass range,
implying that $\sim 120$ stars have masses $M\geq 20\msun$, consistent with
observations -- 90 massive WR/O stars have been observed with $\sim 75\%$
spectroscopic completeness \citep{bartko09}, and these typically have masses
$m>20\msun$ \citep{pau06}. The corresponding disc mass is
$6.3\times10^3\msun$.

\subsection{Normal modes}\label{sec:Modes}

As discussed above, the evolution of the disc is most easily described
in terms of its normal modes. We generate 1000 Monte Carlo
realizations of the disc (stellar masses and semi-major axes).  For
each realization, we calculate the matrices $\bfssA$ and $\bfssO$
(eqs.\ \ref{eq:adef} and \ref{eq:ortho}), as well as the eigenvectors
(the columns of $\bfssO$). As explained in \S\ref{sec:growth}, the
evolution of the disc is determined by the relation between three
characteristic time-scales: the inverse normal mode frequencies
$\Lambda_i^{-1}$, the coherence time $\tau$ of torques from the
spherical cluster, and the age of the disc $t$.

\begin{figure}
\centering
\includegraphics[width=84mm]{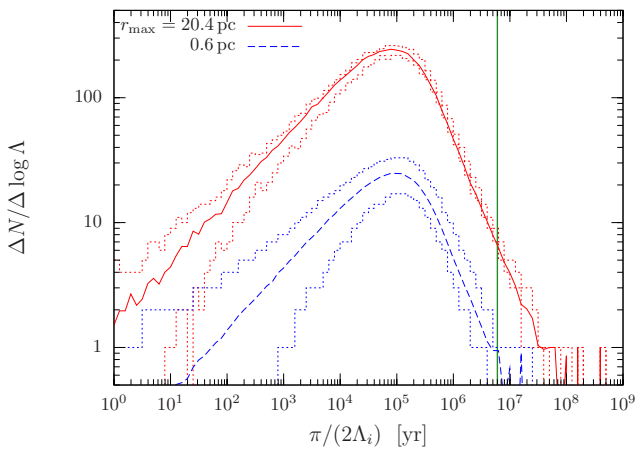}
\caption{Histogram of saturation time-scales $\half \pi/\Lambda_i$ for a disc
  of stars distributed between $1\arcsec$ and $15.5\arcsec$ (dashed blue) or
  extrapolated to $525\arcsec$ (solid red) with surface density $\Sigma(r)\propto
  r^{-1.4}$.  The stellar masses are chosen from equation (\ref{eq:imf}) with
  $\alpha=-0.45$, $m_{\rm max}=30\msun$,
 and $m_{\rm min}=1\msun$; the total
  number of stars is 500 and 5000 in the two cases. Dotted lines show the
  $95\%$ confidence interval. The peak of the distribution
  corresponds to eigenmodes with average wavelengths comparable to
  the average distance between neighbouring stars.
  The age of the disc(s) in the Galactic center is
  marked by a vertical green line. }
\label{fig:Lambdahist}
\end{figure}

Figure~\ref{fig:Lambdahist} shows the distribution of the saturation
time $t_{{\rm s},i}=\half\pi/\Lambda_i$, the characteristic time-scale
at which the amplitude of a mode subjected to a fixed torque stops
growing and begins to oscillate (see \S\ref{sec:coherent}).  The
bottom (blue) curves show the mean distribution and the $95\%$
confidence interval for our standard disc model.  The figure shows
that all but a few modes satisfy $t_{{\rm s},i}<6\Myr$, and so are
already saturated at the current age of the disc.

\begin{figure*}
\centering
\includegraphics[width=58mm]{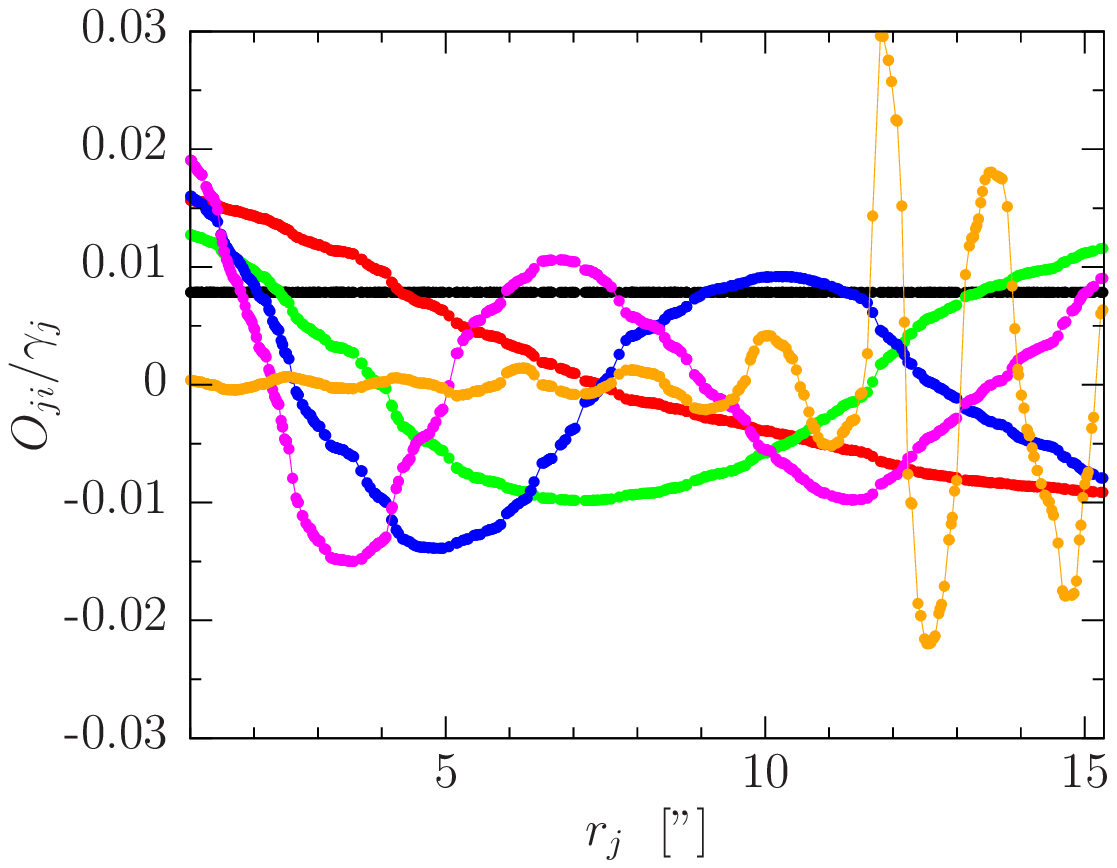}
\includegraphics[width=58mm]{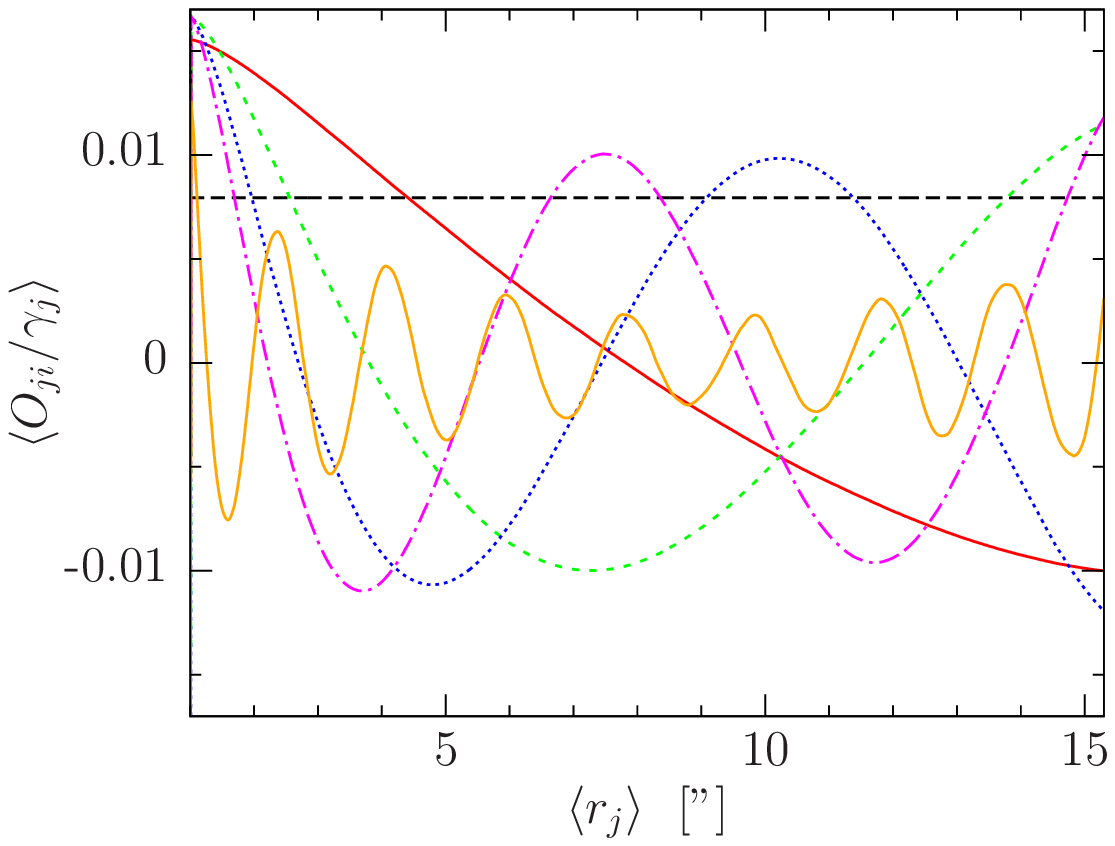}
\includegraphics[width=58mm]{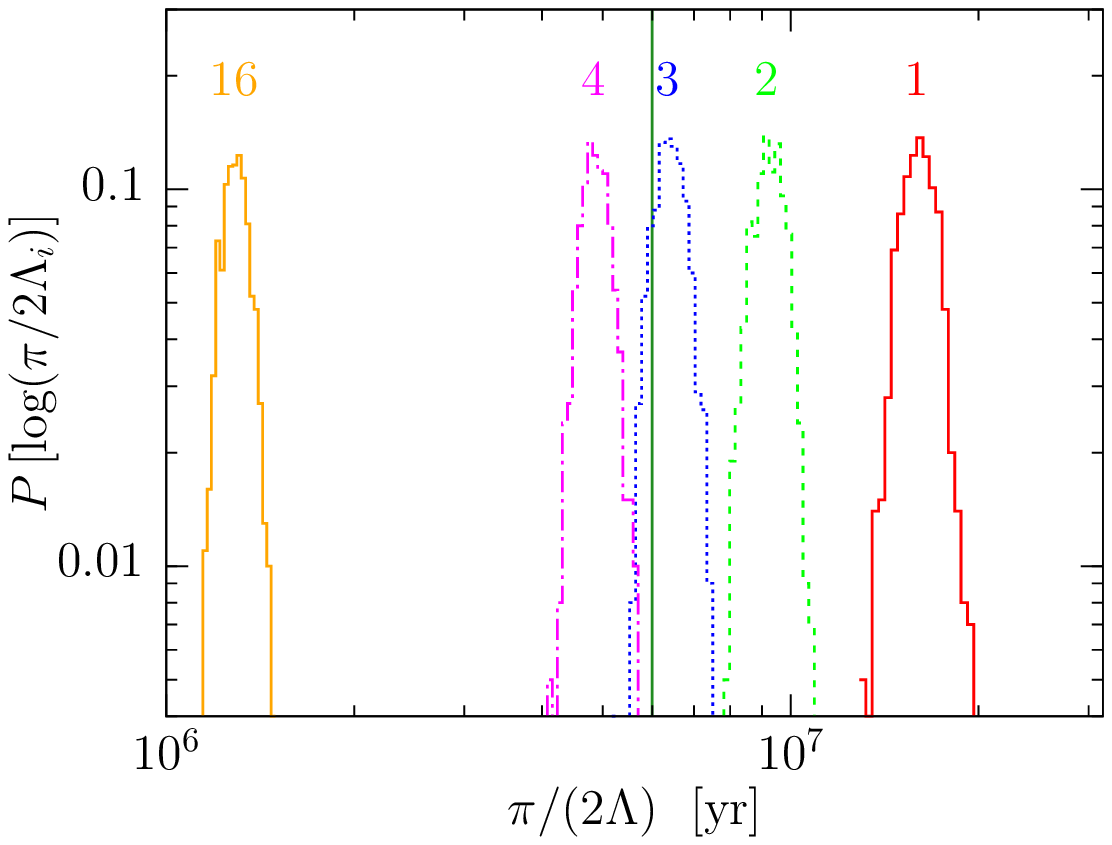}
\caption{ Normal modes of the stellar disc $i=0,1,2,3,4,$ and $16$
  (ordered by increasing frequency) for a randomly chosen realization
  {\it (left)}, and averaged over 1000 realizations of the disc {\it
    (middle)}. The vertical axis is proportional to the fractional
  height of the mode or the inclination angle at fixed azimuth.
These modes of the stellar disc oscillate independently in the absence of the cluster.
{\it Right:} Probability
  distribution of the saturation times for the same modes.  A vertical
  green line shows the disc age $t=6\Myr$. The models contain 500 disc
  stars with masses between $1\msun$ and $30\msun$ and radii between
  $1\arcsec$ and $30\arcsec$.
}
\label{fig:Modes}
\end{figure*}

Some sample normal modes are shown in Figure \ref{fig:Modes}. The
height of the orbit of star $j$ above the reference plane at
azimuth $\phi$ is
$z_j=r_jI_j\sin(\phi-\Omega_j)=-\gamma_j^{-1}r_j(q_j\cos\phi
+p_j\sin\phi)$. Thus if only normal mode $i$ is present, with
amplitude $(Q_i,P_i)$, the height is $z_j=-r_j(Q_j\cos\phi
+P_j\sin\phi )O_{ji}/\gamma_j$.  The figure plots $O_{ji}/\gamma_j$
which is proportional to the fractional height $z_j/r_j$ at fixed
azimuth. The left panel shows a single realization of our standard
disc model and the centre panel shows an average over 1000
realizations. The modes are
ordered by increasing frequency; mode $i=0$ (the black horizontal line in the
figure) has frequency $\Lambda_0=0$ (eq.\ \ref{eq:xxx}), and corresponds to a
uniform tilt of the disc.
\begin{figure*}
\centering
\includegraphics[width=58mm]{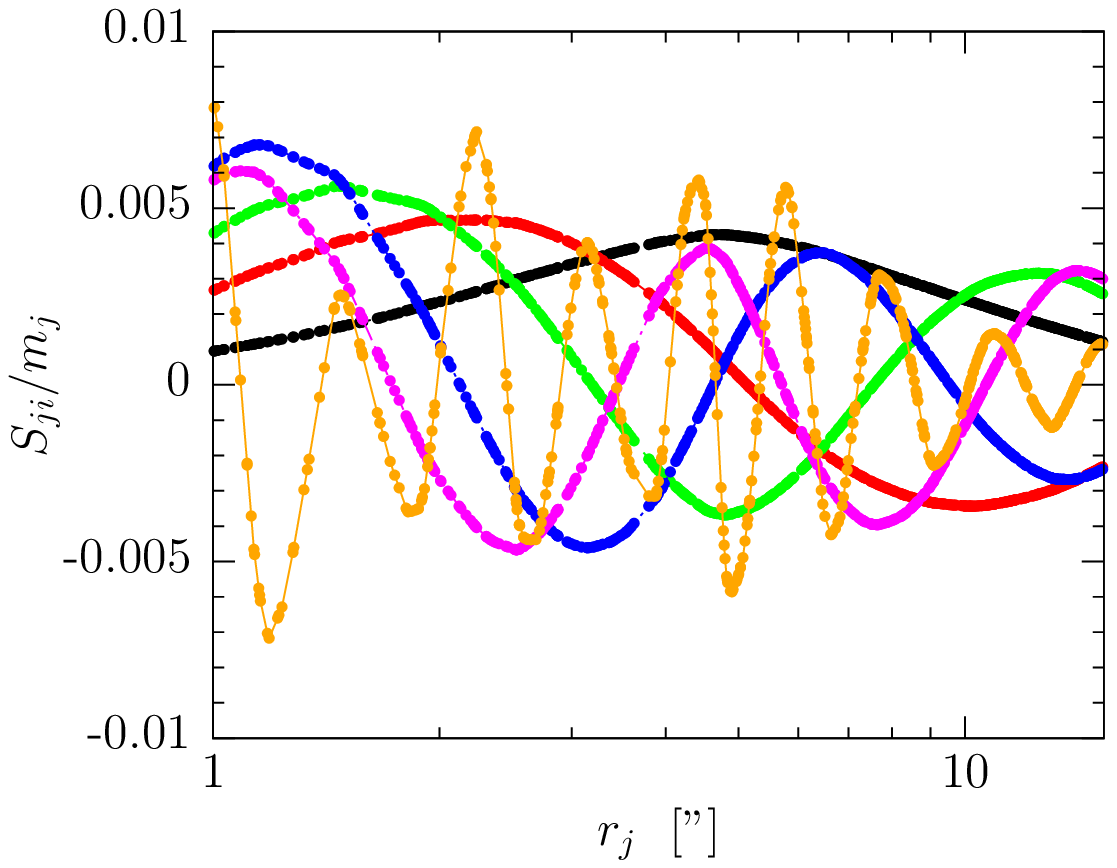}
\includegraphics[width=58mm]{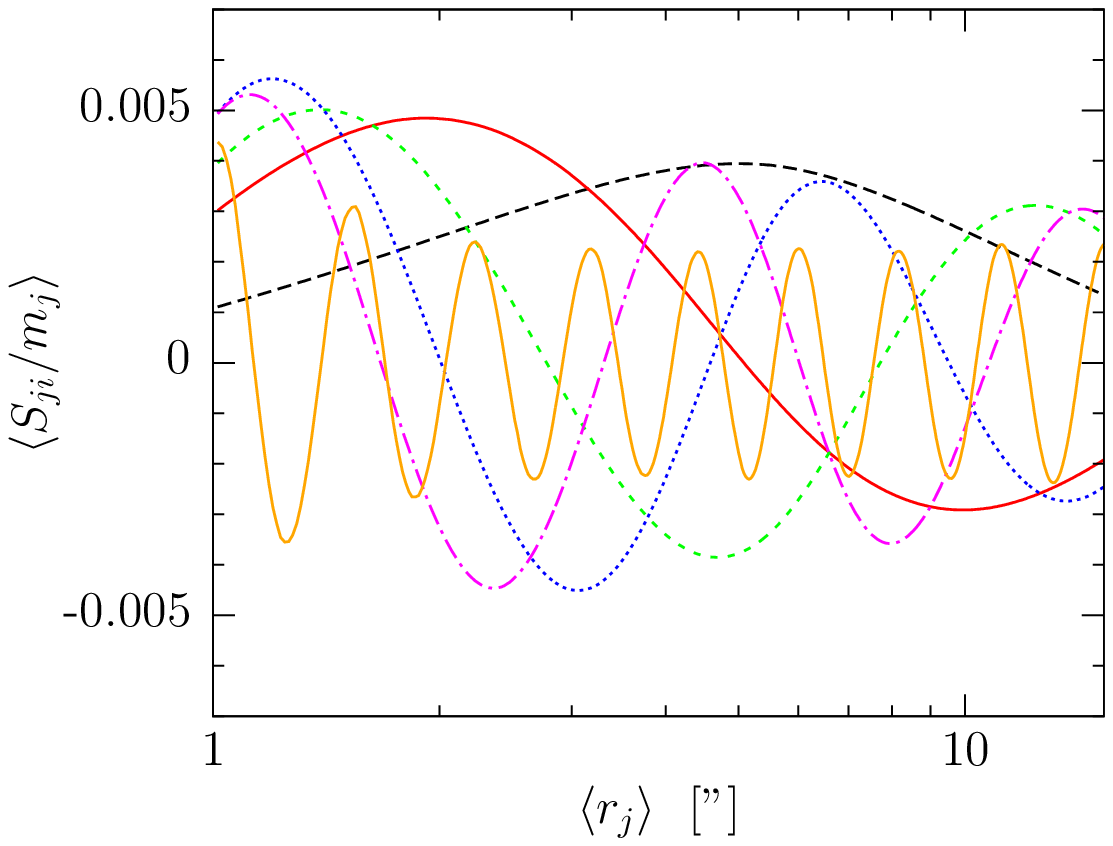}
\includegraphics[width=58mm]{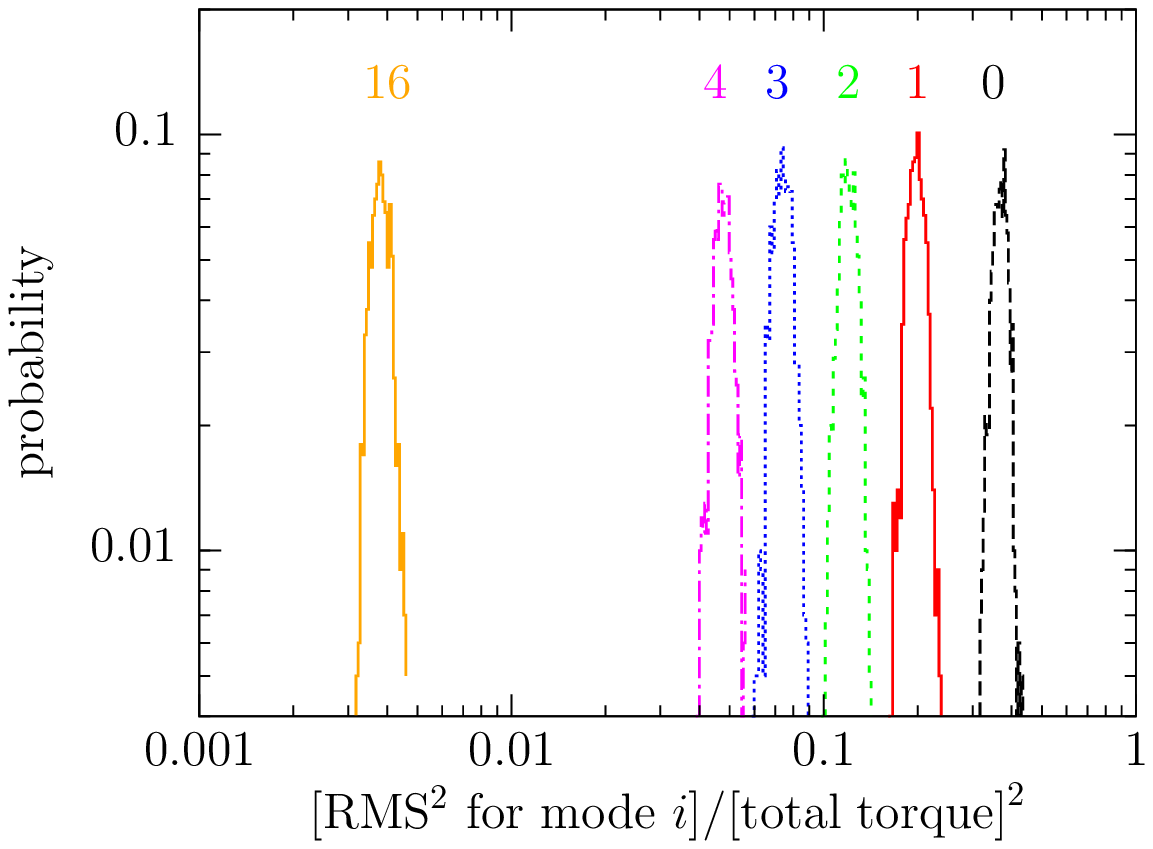}
\caption{ Driving modes of the spherical cluster, i.e., the
  eigenvectors of the spatial correlation function of the torques on
  the disc, $\langle T_{xi}T_{xj}\rangle$. The figure shows modes $i=0,1,2,3,4,$ and $16$
  (ordered by decreasing significance) for a randomly chosen realization
  {\it (left)}, and averaged over 1000 realizations of the disc {\it
    (middle)}.
  The models contain the same stars as in Fig.~\ref{fig:Modes}.
  {\it Right:} Probability distribution of the mean squared amplitude
  for the same modes.
}
\label{fig:ExcitationModes}
\end{figure*}

The low-frequency modes are rather smooth, and their shapes are approximately
the same in different realizations of the disc -- the mode shapes are not
sensitive to the large random variations in the masses of individual stars
($1\msun\leq m_j\leq 30\msun$). The long-wavelength modes are
well represented by the average waveform, while shorter wavelength modes in a
single realization deviate more strongly from the mean
(compare the orange lines, $i=16$, in the left and centre panels of Figure
\ref{fig:Modes}). The right panel shows the probability
distribution for the saturation times $\half\pi/\Lambda_i$ for various
realizations of the disc.
The probability distribution is sharply peaked for $\Lambda_i$ (for each
fixed $i$) with a FWHM of about $20\%$.  Therefore, $\Lambda_i$ can be
predicted from the surface-density distribution without knowing the locations
and masses of disc stars.
The normal modes $i=0,1,2$ are still growing ($t_{{\rm s},i}>6\Myr$), $i=3$ is
just around saturation, and all other modes are saturated.

These considerations and the theory presented in \S\ref{sec:coherent} allow us
to make general remarks on the expected warping of the disc. Recall that modes
in the oscillating phase -- with either the coherence time or the disc age less
than the saturation time $t_{{\rm s},i}=\half\pi/\Lambda_i$ -- are suppressed in
amplitude relative to lower frequency modes by $\Lambda_i^{-1}$
(see eq.~[\ref{eq:correlatedgrowth}] and \S\ref{sec:warp} below).
Therefore if the external torques on the disc stars do not depend too strongly on
radius, the shape of the disc is expected to be dominated by long-wavelength
modes.

There are presently few observational constraints on the maximum radial extent
of the disc, and its evolution would be different if it extended to larger radii. As
an example, the top (red) curves in Figure \ref{fig:Lambdahist} show the
histogram of saturation times for a hypothetical disc extrapolated to
$20.4\pc$ with the same power-law surface density distribution.  In
this case, there are many more modes still in the growing phase at the current
age of $6\Myr$.
However, the characteristic wavelengths of these additional modes are
larger than $0.6\pc$, and the saturation timescales of the smaller wavelength
modes are not very different from the case in which the disc is truncated
at $0.6\pc$. Therefore we expect that our predictions of the shape of
the warped disc at radii $\lesssim 0.5\pc$ are robust, and independent
of whether or not there is an outer disc.

\subsection{Spherical cluster}

We model the torques on the disc from the star cluster as follows: (i)
As described above, since the torque distribution is a Gaussian random
process, we may assume that all cluster stars have the effective mass
$m_2$, as defined following equation (\ref{eq:trelax}); we typically
assume $m_2=10\msun$. (ii) As described above, for the sake of
simplicity we assume that the cluster stars are on circular orbits,
and generate a distribution of orbital radii consistent with the mass
distribution $M(r)$ in equation (\ref{eq:mass}) out to a radius of
$2\pc$, sufficiently far outside the disc(s) that the perturbations
from larger radii should be small.  (iii) We assign nodes
$\Omega_\beta$ uniformly random between 0 and $2\pi$ and inclinations
$I_\beta$ so that $\cos I_\beta$ is uniformly random between $-1$ and
$+1$, consistent with an isotropic distribution of orbits. (iv) We
expand the torque (\ref{eq:torque}) to order $2\ell=10$. (iv) We
assume a disc age of $6\Myr$.

Alternatively, random realizations of cluster torques may be generated
by sampling an $N$-dimensional Gaussian distribution (where $N$ is the number of
disc stars) with a covariance matrix $\langle T_{xi} T_{xj}\rangle$
given by equations (\ref{eq:sigmaij2})--(\ref{eq:sigmaij3}).
We can decompose the torques
into independent driving modes, corresponding to the eigenvectors of
$\langle T_{xi} T_{xj}\rangle$. These driving modes, shown in
Figure \ref{fig:ExcitationModes}, are a property of the spherical
cluster of old stars and hence are distinct from the
normal modes of the disc shown in Figure \ref{fig:Modes}.
We find that the corresponding eigenvalues,
which correspond to the mean squared amplitude of the particular driving mode,
span a vast range, some 6 orders of magnitude. Remarkably, the five
longest wavelength driving modes typically contribute 90\% of the total torque.
Therefore, the cluster excites the disc predominantly through a few
long-wavelength driving modes.

As described at the end of \S\ref{sec:incoh}, the coherence time of the
torques from the cluster can be shorter than the disc age depending on the
radius.  The two are equal at a radius around $r_{\tau}\sim0.3\pc (m_2/10\msun)^{0.56}$.
Nevertheless, to avoid excessive complication -- both conceptual and
numerical -- in our simulations we shall assume that the coherence time is long
compared to the disc age, so that the perturbing forces $f_{qi}(t)$, $f_{pi}(t)$ can be
regarded as time-independent and the formulae in \S\ref{sec:coherent} can be
used to calculate the disc evolution. This simplification has a smaller impact
on the evolution of long wavelength modes $\lambda\gsim r_{\tau}$,
while it may be important for short wavelength modes close to the
center.
To account properly for the temporal evolution of the torques from the
cluster would require solving simultaneously for the secular evolution
of both the cluster and the disc stars (see \S\ref{sec:discussion}).

\subsection{Warped disc}\label{sec:warp}
\begin{figure}
\centering
\includegraphics[width=84mm]{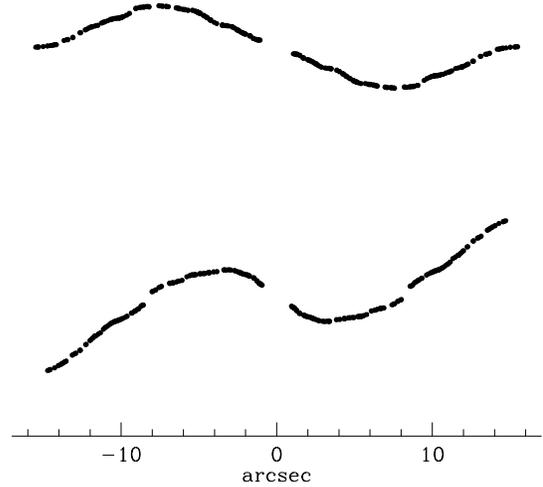}
\caption{Two orthogonal cross-sections of a disc subjected to vector resonant
  relaxation. The parameters of the disc and the stellar cluster in which it
  is embedded are described in \S\ref{sec:monte}. The disc was initially
  horizontal. The calculation is based on the assumption that the inclinations
  are small and the results presented here are large enough that this
  approximation is not quantitatively accurate. Only stars with mass
  $>20\msun$ are shown but the stars of lower mass have the same thin, warped
  distribution. The simulation assumes that the disc age $t=6\Myr$ and the
  effective mass in the stellar cluster is $m_2=10\msun$; the inclinations
  scale as $m_2^{1/2}$.}
\label{fig:warp}
\end{figure}

The results of a typical simulation are shown in Figure \ref{fig:warp}, which
plots two orthogonal cross sections through the disc. The disc remains thin,
but exhibits a substantial warp with a structure dominated by the
zero-frequency tilt mode and the three longest wavelength normal modes of the
disc $j=1,2,3$.  We conclude that vector resonant relaxation can excite a
coherent warp in an initially thin, flat disc.

\begin{figure}
\centering
\includegraphics[width=84mm]{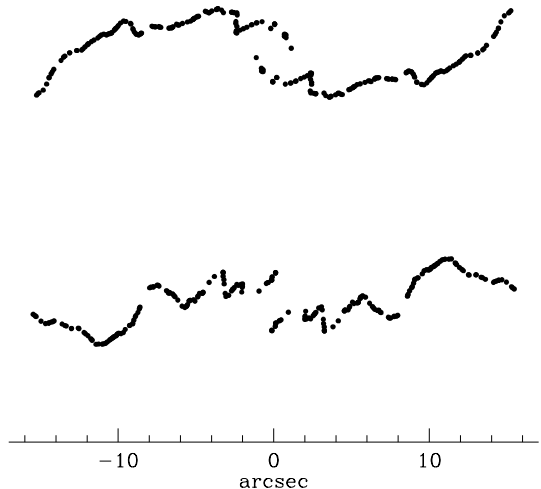}
\caption{Identical to Figure \ref{fig:warp}, except that the gravitational
  interactions among the disc stars have been turned off. The
  disc-star inclinations are much more irregular at small radii.
}
\label{fig:warpone}
\end{figure}

In two-body relaxation the perturbing forces from a surrounding cluster of
stars usually thicken a disc, rather than warping it. For example, the
observed thickness of the Galactic disc in the solar neighborhood has been
used to constrain the effective mass of the objects comprising the Galaxy's
dark-matter halo \citep[e.g.,][]{lo85}. Warping, rather than thickening,
occurs in this case for two main reasons:

(i) Vector resonant relaxation arises through the torque from a stellar orbit
after averaging over mean anomaly and argument of pericentre, and this
averaged torque has much less small-scale power than the forces exerted in the
passage of a nearby star (cf.\ Fig.\ \ref{fig:ExcitationModes}). More
quantitatively, consider two nearby stars on circular orbits with semi-major
axes $a$ and $a+\Delta a$, and relative inclination $I\lesssim \Delta a/a$;
torques from stars at larger relative inclinations are smaller and can be
neglected for this argument. The time-averaged or secular torque between the
two orbits is $T\sim Gm^2/\Delta a$. For a radial density profile
$n\sim n_0 r^{-\gamma}$,
the number of stars in this semi-major axis and inclination range is
$\Delta N \sim n_0 a^{2-\gamma} I^2 \Delta a \sim n_0 a^{-\gamma} (\Delta a)^3$.
The total stochastic torque generated by these stars is then
$(\Delta N)^{1/2}T \propto n_0^{1/2} Gm^2 (\Delta a)^{1/2}$.
This is an increasing function of $\Delta a$, showing that the secular torque
on the disc is dominated by large-scale fluctuations. A nonzero eccentricity for
either the cluster or disk stars further reduces the small-scale power.

(ii) As suggested by equations (\ref{eq:eqmotb}), the excitation of normal
modes in the disc is reduced if the coherence time of the torque exceeds the
inverse frequency of the normal mode, and the coherence
time for vector resonant relaxation is longer than the inverse frequencies
associated with small-scale normal modes in the disc. To illustrate the
importance of this effect, Figure \ref{fig:warpone} shows the
results of a simulation identical to that leading to Figure \ref{fig:warp},
except that the masses of the disc stars have been reduced to nearly zero.
Reducing the disc-star masses decreases all oscillation
frequencies $\Lambda_i$, since these are proportional to mass in
Laplace--Lagrange theory. In the limit of near-zero mass, the
interaction between the disc stars becomes negligible
as described in Sec.~\ref{sec:initialresponse} and
$\langle Q_i Q_j \rangle$ becomes independent of $\Lambda_i$.  Figure
\ref{fig:warpone} shows that in this limit the warp
becomes much more irregular and less spatially coherent, because both
large-scale and small-scale modes are excited.

\begin{figure}
\centering
\includegraphics[width=84mm]{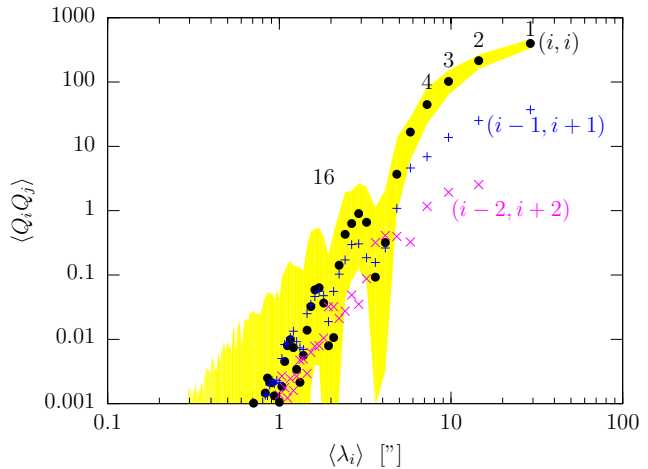}
\caption{The typical amplitudes of normal modes (arbitrary scale), as a function of their average
  wavelength.
  The circles mark the mean of the squared amplitudes of normal modes
  for a random realization of the disc for each $i$, and the 95\% confidence interval is shaded.
  Plusses and crosses show the cross--correlation between various normal modes
  for comparison. In this case,  the distribution is symmetric around 0,
  and the symbols are placed at the boundary of the 95\%  confidence interval.
  The modes $i=1,2,3,4,$ and 16 shown in Fig.~\ref{fig:Modes} are labelled. The torques
  are assumed to be constant during the $6\Myr$ lifetime of the disc with
  an amplitude corresponding to the RMS torque  (eq.~\ref{eq:sigmaij3}).
  The high--frequency, short--wavelength normal modes are suppressed.}
\label{fig:growth}
\end{figure}
Figure~\ref{fig:growth} shows the normal mode power and cross-correlation
as a function of the average wavelength, $\lambda_i = 2 (r_{\max}-r_{\min})/i$.
The $y$--axis shows $\langle Q_{i+k} Q_{i-k}\rangle$ for $k=0$, 1, and 2,
where $Q_i$ is the amplitude of the $i^{\rm th}$
normal mode (see Fig.~\ref{fig:Modes}).
The circles show the mean of $Q_i^2$ and the shaded region
shows the 95\% confidence interval for a random realization of the disc.
In contrast, the cross correlation of normal modes has zero mean,
for these elements the symbols correspond to the upper boundary of the 95\% interval.
The calculations are based on equations (\ref{eq:correlatedgrowth}) and
(\ref{eq:sigmaij3}). The long-wavelength modes dominate the disc, and modes
with wavelength $\lesssim 1\arcsec$ are suppressed by some five orders of
magnitude. Note that the predicted amplitude of the longest wavelength modes
($i\leq 6$) exhibits only a small scatter between various realizations
of the disc. Although the different normal modes are not independent,
the cross correlation between modes with increasingly different average
wavelengths are clearly more and more suppressed.
The RMS warping of the disc then follows from eq.~(\ref{eq:IRMS}).
Figure~\ref{fig:growth} shows that a coherent torque tilts and warps the
disc in a way that its shape is reminiscent of only a few long wavelength
modes.

\section{Discussion}
\label{sec:discussion}

The massive young stars observed in the Galactic centre at radii $\sim
0.05$--0.5 pc from the central black hole may have formed in a thin, flat
disc. Using the best available estimates for the mass, age, and other
properties of the disc and the cluster of old stars in which it is embedded,
we have shown that the disc naturally and inevitably develops a strong
warp through vector resonant relaxation with the cluster. We suggest
that this mechanism explains the large ($\sim 60^\circ$) warp that is observed
in the disc.

We have modelled the gravitational torques among the disc stars and between the
disc stars and the surrounding old cluster. Our models are based on
Laplace--Lagrange theory, in which the secular gravitational interactions of
the disc stars are mathematically equivalent to the interactions in a system
of masses connected by springs. Laplace--Lagrange theory assumes
that the relative inclinations are small,
$\Delta_{jk}\equiv|a_jI_j-a_kI_k|/|a_j-a_k|\ll1$.  For the simulation shown in
Figure \ref{fig:warp}, the maximum value of $\Delta_{jk}$ was 2.8, indicating
that Laplace--Lagrange theory has been extended beyond its domain of
applicability but should still be qualitatively correct.

In Laplace--Lagrange theory, the disc dynamics can be decomposed into normal
modes. The eigenfrequencies of the normal modes in our models span a vast
range of time-scales (cf.\ Fig.\ \ref{fig:Lambdahist}). The high-frequency
normal modes are an unphysical consequence of the assumption that the disc
orbits are precisely circular and that the inclinations are small compared to
the separation between stars; in fact the highest frequencies correspond to
modes involving only two adjacent stars. However, our results should not be
affected by this shortcoming since the high-frequency modes are not excited
efficiently by torques from the cluster stars (cf.\ Fig.\ \ref{fig:growth}).

The frequencies depend monotonically on the wavelength of the mode:
oscillation times $\pi/\Lambda$ of about $13 \Myr$ correspond to
wavelengths of $\sim 0.4\pc$ ($10\arcsec$) (cf.\ blue curve in Fig. 4);
faster or slower modes have smaller and longer wavelengths,
respectively.  The frequencies and the shapes of the slow normal modes
are insensitive to the details of the stellar mass function so long as
the large--scale surface density of the disc is fixed.

The growth of a given mode saturates after about
a half oscillation period. The low-frequency, long-wavelength modes
are excited to much larger amplitudes than high-frequency, short-wavelength
modes, both because the secular torques couple more strongly to
long-wavelength modes, and because the low-frequency modes saturate after a
longer time (an equivalent statement is that the amplitudes of the
high-frequency modes are adiabatic invariants).

The most important free parameter in our models is the effective mass
$m_2=\langle m^2\rangle/\langle m\rangle$ of the stars and other
irregularities (e.g., clusters, gas clouds, etc.) in the cluster; the
simulation in Figure \ref{fig:warp} assumes $m_2=10\msun$, which is likely to
be an underestimate.
A larger value of $m_2$ increases the warp amplitude proportional to $m_2^{1/2}$.

For simplicity our simulations are done using time-independent torques from
a cluster.  For a spherical cluster, this approximation is valid if the disc
age is less than the vector resonant relaxation time, since this is
approximately the same as the coherence time. A rough measure of the validity
of this approximation is given by Figure \ref{fig:warpone}, which shows the
evolution of the orientations of the stars in a disc of zero mass: since the
disc self-gravity has been turned off in this figure, the stars behave in
the same manner as random stars in the spherical cluster, so the approximation
that the perturbing forces from the cluster are constant is valid if the
inclinations of the stars in this Figure are small. Evidently the approximation of constant
forces is good for the outer parts of the cluster, where the relaxation time
is relatively long, and suspect in the inner parts.

Both of our major approximations -- Laplace--Lagrange theory and constant perturbing
forces -- are more accurate if the effective mass is smaller than the assumed
value $m_2=10\msun$ and worse if it is larger.

We also assumed that the cluster is spherical on average (i.e., apart
from Poisson fluctuations due to individual stars), that the
orientations of cluster star orbits are uncorrelated,  that there
is no back-reaction from the disc on the cluster, that the cluster
stars are on circular orbits, and that the cluster mass profile $M(r)$
(eq.\ \ref{eq:mass}) can be determined from the luminosity distribution $L(r)$ assuming constant
mass-to-light ratio. Under these
approximations, the fractional deviation from sphericity at radius
$r\gtrsim 0.2\pc$ is roughly $\sqrt{m_2/M(r)} = 0.009
\sqrt{m_2/10\msun} (r/0.1\pc)^{-0.63}$. The deviation from
sphericity may be larger if the cluster is flattened due to rotation;
however, \citet{tri08} estimate the rotation speed of the cluster to
be $v_{\rm rot}(r)=(3.6\pm 0.8)\kms (r/0.1\pc)$ within 1\,pc.  The
rotational flattening of the cluster is then roughly $ v_{\rm rot}^2 /
\sigma^2 \sim 2\times 10^{-4} (r/0.1\pc)^3$, smaller than the
stochastic flattening for $r\lesssim 0.3\pc$. The flattening of the
overall mass distribution due to the disc is comparable to the
stochastic flattening (see discussion following eq.\ \ref{eq:tresv})
so the neglect of this contribution probably does not seriously
invalidate our results. The justification for neglecting the
back-reaction of the disc is described at the end of
\S\ref{sec:torquedist}. The assumption that the orientations of
cluster star orbits are uncorrelated is harder to justify, and to do
so will require a self-consistent simulation of both the disc and
cluster stars (see the discussion at the end of this section).
The assumption that the stars in the cluster are on circular orbits is
unrealistic -- the scalar resonant relaxation time is much less than the age of
the Galaxy so we expect the eccentricity distribution to be $dn\propto
e\,de$. This defect will be removed in future calculations.

\cite{ba09} have suggested that warps in accretion discs surrounding black
holes in the centres of galaxies -- in particular, the $\sim 10^\circ$ warp in
the maser disc in NGC 4258 -- may arise from resonant relaxation with the
surrounding stellar cluster. In accretion discs the warp dynamics may
also be affected by gas pressure, viscous dissipation, and radial
transport of mass and angular momentum \citep{pringle92,lp07},
as well as opacity and radiation pressure
in radiatively efficient discs \citep{petterson77,mbp96}.

The treatment in the present paper does not address several important issues:

\begin{itemize}

\item What are the characteristics of the disc after vector resonant
  relaxation excites a large warp -- in particular, large enough that the
  Laplace--Lagrange treatment is invalid?

\item Can vector resonant relaxation create structures that resemble the
  second, `counter-clockwise' disc seen at the Galactic centre? A mechanism
  to form both the clockwise and counter-clockwise discs from a single thin,
  flat precursor would explain why stars in both discs have the same age.

\item The WR/O stars in the disc(s) are much more massive than the old cluster
  stars. What is the analog of dynamical friction for vector resonant
  relaxation and what role does it play in determining the disc structure?

\end{itemize}

These questions can be addressed most effectively by `N-ring' simulations that
follow the evolution of a set of $N$ axisymmetric rings, each representing the
smeared-out mass density in a Keplerian orbit after averaging over mean
anomaly and argument of pericentre. Each ring exerts a torque on every other
ring, and the simulation follows the precession of the rings in response to
these torques. We have written a code to carry out these simulations and are
currently using it to follow the evolution of stellar systems that resemble
the central parsec of the Galaxy.

\bigskip

We thank Yuri Levin for useful discussions and the anonymous referee for
suggestions that substantially improved the paper.
This research was supported in part by NASA grant NNX08AH24G
and NSF grant AST-0807432.  B.K. acknowledges support by NASA through
Einstein Postdoctoral Fellowship grant number PF9-00063 awarded by the
Chandra X-ray Center, which is operated by the Smithsonian
Astrophysical Observatory for NASA under contract NAS8-03060, and
partial support by OTKA grant 68228.

\onecolumn

\appendix
\section{The microcanonical ensemble in Laplace--Lagrange theory}

We have argued in \S\ref{sec:timescale} that vector resonant relaxation among
the stars of the Galactic-centre disc(s)  is unimportant. However, there
is a wide range of disc ages and environments for which internal vector
resonant relaxation may be the dominant process of dynamical relaxation. A
particularly simple case is an isolated disc with age much larger than the
vector resonant relaxation time. Isolated discs conserve total energy and
angular momentum; moreover, if vector resonant relaxation is the only important
dynamical process the semi-major axis and eccentricity of each star is
conserved. According to the usual principles of statistical mechanics, the
probability distribution of such discs (the microcanonical ensemble) should be
uniform in the manifold of phase space that is determined by these conserved
quantities.

For sufficiently small inclinations, the Hamiltonian of the disc is given by
the Laplace--Lagrange form (\ref{eq:hamsec}). Since this Hamiltonian is
quadratic in the coordinates and momenta, the equations of motion are linear
and therefore integrable. Thus in principle an isolated disc with
sufficiently small eccentricities and inclinations exhibits no
resonant relaxation. Our assumption is that higher-order terms that we have
neglected in the Hamiltonian lead to relaxation to an equilibrium state that
can be approximately described using the statistical mechanics of the
quadratic Hamiltonian -- just as occasional collisions in an ideal gas lead to
a thermal equilibrium that can be described using the quadratic Hamiltonian
$H=\half p^2/m$.

The Hamiltonian $H$ and the angular-momentum deficit $Z$ (eq.\ \ref{eq:zdef})
are both conserved. An important combination of these is
\begin{equation}
  \Gamma(\bfq,\bfp)\equiv \frac{H(\bfq,\bfp)}{
  Z(\bfq,\bfp)}=\frac{\bfp^{\rm T}\bfssA\bfp+\bfq^{\rm
      T}\bfssA\bfq}{\bfp^{\rm T}\bfp+\bfq^{\rm T}\bfq}.
\label{eq:omdef}
\end{equation}
Since $\Gamma$ has units of inverse time we call it the frequency parameter of
the disc\footnote{Note that $\Gamma$ used here is not related to $\Gamma_{n m}(t)$
denoting the correlation function in the main text.}.

We can re-write the last of these expressions as
\begin{equation}
\Gamma(\bfP,\bfQ)=\frac{\sum_{i=0}^{N-1}\Lambda_i(P_i^2+Q_i^2)}{
           \sum_{i=0}^{N-1} (P_i^2+Q_i^2)},
\label{eq:omdefa}
\end{equation}
which shows that $\Gamma$ can only span a limited range of values,
from $\Lambda_{\rm min}$ to $\Lambda_{\rm max}$ where these are the minimum
and maximum eigenvalues of $\bfssA$.

We may also assume that the conserved $x$ and $y$ components of the angular
momentum $L_x$, $L_y$ are both zero, since this can be ensured by choosing the
$z$-axis of the coordinate system to be parallel to the total angular-momentum
vector. Then from equation (\ref{eq:xxx}) $P_0=Q_0=0$.

To analyse the properties of the microcanonical ensemble, we first
find the density of states $\omega_N(E,C)$, defined so that
$\omega_N(E,C)dEdC$ is the volume in $2(N-1)$-dimensional phase space
in which the energy and angular-momentum deficit are in the ranges
$(E,E+dE)$ and $(C,C+dC)$. Thus\footnote{The system we are studying
  resembles the celebrated spherical model for a ferromagnet
  \citep{bk52} and much of our analysis is borrowed from the
  literature on that problem.}

\begin{align}
\omega_N(E,C) =&\int \prod_{i=i}^{N-1} dP_idQ_i\,\delta\big[E-
H(\bfP,\bfQ)\big]\delta\big[C-Z(\bfP,\bfQ)\big]\notag \\ =&
\int \prod_{i=1}^{N-1} dP_idQ_i\,\delta\big(E-\bfP^{\rm T}\bfssLambda\bfP+\bfQ^{\rm
  T}\bfssLambda\bfQ\big)\,\delta\big(C-\bfP^{\rm T}\bfP-\bfQ^{\rm T}\bfQ\big).
\end{align}
We use the identity
\begin{equation}
\delta(x)=\frac{1}{2\pi i}\int_{-i\infty}^{i\infty}dt\, e^{tx}.
\end{equation}
Then
\begin{align}
\omega_N(E,C)=&-\frac{1}{4\pi^2}\int \prod_{i=1}^{N-1}dP_idQ_i \int_{-i\infty}^{i\infty}\!\!dt\,
\int_{-i\infty}^{i\infty}\!\!dt' \exp\big[tE+t'C-t(\bfP^{\rm T}\bfssLambda\bfP+\bfQ^{\rm
  T}\bfssLambda\bfQ)-t'(\bfP^{\rm T}\bfP+\bfQ^{\rm
  T}\bfQ)\big]\notag \\
=&-\frac{1}{4\pi^2}\int \prod_{i=1}^{N-1}dP_idQ_i \int_{-i\infty}^{i\infty}\!\!dt\,
e^{tE}\int_{-i\infty}^{i\infty}\!\!dt'\, e^{t'C}
\exp\Big[-\sum_{i=1}^{N-1}(P_i^2+Q_i^2)(t\Lambda_i+t')\Big].
\end{align}
The next step is to exchange the order of integration, but this cannot be done
immediately as $\int dPdQ\,\exp[-(P^2+Q^2)(t\Lambda+t')]$ is not absolutely convergent
for imaginary $t$ and $t'$. However, using the identity
$1=\exp(uC)\exp[-u\sum_{i=1}^{N-1}(P_i^2+Q_i^2)]$ (recall $P_0=Q_0=0$) we can
rewrite the double integral as
\begin{equation}
\omega_N(E,C)=-\frac{e^{uC}}{4\pi^2}\int \prod_{i=1}^{N-1}dP_idQ_i\int_{-i\infty}^{i\infty}\!\!dt\,
e^{tE}\int_{-i\infty}^{i\infty}\!\!dt'\, e^{t'C} \exp\Big[-\sum_{i=i}^{N-1}(P_i^2+Q_i^2)(t\Lambda_i+t'+u)\Big],
\label{eq:multint}
\end{equation}
and the integral over $dP_idQ_i$ is now absolutely convergent for $u>0$. The order
of integration can now be exchanged so that the $2(N-1)$-dimensional integral over
phase space can be done:
\begin{equation}
\omega_N(E,C)=-\frac{\pi^{N-3}}{4}\int_{-i\infty}^{i\infty}\!\!dt\,
e^{tE}\int_{u-i\infty}^{u+i\infty}\!\!ds\,e^{sC}\prod_{i=1}^{N-1}(s+t\Lambda_i)^{-1}
=-\frac{\pi^{N-3}}{ 4}\int_{-i\infty}^{i\infty}\!\!dt\int_{u-i\infty}^{u+i\infty}\!\!ds\,
\exp\Big[tE+sC-\sum_{i=1}^{N-1}\log(s+t\Lambda_i)\Big],
\label{eq:gdef}
\end{equation}
where $s\equiv u+t'$.

For $N\gg1$ we can evaluate this integral by the method of stationary phase.
Denoting the quantity in square brackets as $g(s,t)$, the dominant
contribution to the integral comes from near the points $(s_0,t_0)$ at
which $\p g/\p s=\p g/\p t=0$, that is,
\begin{equation}
C=\sum_{i=1}^{N-1}\frac{1}{s_0+t_0\Lambda_i}, \quad
E=\sum_{i=1}^{N-1}\frac{\Lambda_i}{s_0+t_0\Lambda_i}.
\label{eq:eldef}
\end{equation}
We now show that $s_0$ and $t_0$ are real. Since $E$, $C$, $u$, and
$\Lambda_i$ are real, the imaginary part of these equations reads
\begin{equation}
w_0\Im(s_0)+w_1\Im(t_0)=0,\quad w_1\Im(s_0)+w_2\Im(t_0)=0,
\end{equation}
where
\begin{equation}
w_n\equiv \sum_{i=1}^{N-1}\frac{\Lambda_i^n}{D(\Lambda_i,s_0,t_0)}, \quad
D(\Lambda_i,s_0,t_0)=[\Re(s_0)+\Re(t_0)\Lambda_i]^2+[\Im(s_0)+\Im(t_0)\Lambda_i]^2.
\label{eq:deldef}
\end{equation}
Thus either $\Im(s_0)=\Im(t_0)=0$ or
\begin{equation}
w_0w_2-w_1^2=\half \sum_{i=1}^{N-1}\sum_{j=1}^{N-1}\frac{(\Lambda_i-\Lambda_j)^2}{
D(\Lambda_i,s_0,t_0)D(\Lambda_j,s_0,t_0)}
\label{eq:deldefb}
\end{equation}
must vanish. This condition is only satisfied in the trivial case when all of the
eigenvalues $\Lambda_i$ are equal. Thus $s_0$ and $t_0$ must be real.

Let the real numbers $s_c$ and $t_c$ denote the locations where the
integration contours cross the real axis in the complex $s$ and $t$ planes. In
equation (\ref{eq:gdef}) $s_c=u$ and $t_c=0$. In the method of stationary
phase the contours are deformed to cross the real axes at $s_c=s_0$ and
$t_c=t_0$.  During this deformation $s_c$ and $t_c$ cannot cross the poles at
$s+t\Lambda_i=0$. Each such pole defines a line in the $(s_c,t_c)$ plane and
in its journey from $(u,0)$ to $(s_0,t_0)$ the point $(s_c,t_c)$ cannot cross
any of these lines. This constraint implies that
\begin{equation}
t>-\frac{s}{\Lambda_{\rm min}}\quad\hbox{if $s<0$};\qquad t>
-\frac{s}{\Lambda_{\rm max}}\quad\hbox{if $s\ge0$}
\label{eq:lamlim}
\end{equation}
where $\Lambda_{\rm min}$ is the smallest eigenvalue (other than
$\Lambda_0=0$) and $\Lambda_{\rm max}$ is the largest, i.e.,
$0=\Lambda_0<\Lambda_{\rm min}\le\Lambda_i,\ i=1,\ldots,N\le \Lambda_{\rm
  max}$.

To evaluate the integral (\ref{eq:gdef}) we expand the argument $g(s,t)$ of
the exponential in a Taylor series around $(s_0,t_0)$. We have
\begin{equation}
\frac{\p^2g}{\p s^2}(s_0,t_0)=w_0, \quad \frac{\p^2g}{\p s\p t}(s_0,t_0)=w_1,
\quad
\frac{\p^2g}{\p t^2}(s_0,t_0)=w_2,
{\rm~~~where~~}
w_n = \sum_{i=1}^{N-1}\frac{\Lambda_i^n}{ (s_0+t_0\Lambda_i)^2},
\label{eq:wwwdef}
\end{equation}
which is consistent with (\ref{eq:deldef}) since $s_0$ and $t_0$ are now know
to be real. Then
\begin{equation}
\omega_N(E,C)\simeq  -\frac{\pi^{N-3}}{
  4}\int_{-i\infty}^{i\infty}\!dt\int_{u-i\infty}^{u+i\infty}\!ds
\exp\left[g(s_0,t_0)+\half w_0(s-s_0)^2+w_1(s-s_0)(t-t_0)+\half w_2(t-t_0)^2\right].
\label{eq:gdefa}
\end{equation}
We introduce new integration variables $(v,v')$ by $s=s_0+iv$,
$t=t_0+i(v'-vw_1/w_2)$ and the integral becomes
\begin{align}
\omega_N(E,C)\simeq &\frac{\pi^{N-3}\exp[g(s_0,t_0)]}{
  4}\int_{-\infty+i(s_0-u)}^{\infty+i(s_0-u)}\!dv\exp\bigg(-\frac{w_0w_2-w_1^2}{
  2w_2}v^2\bigg)  \int_{i[t_0+(s_0-u)w_1/w_2]-\infty}^{i[t_0+(s_0-u)w_1/w_2]+\infty}\!dv'\exp[-\half w_2(v')^2]\notag \\
=&\frac{\pi^{N-2}}{ 2\sqrt{w_0w_2-w_1^2}}\exp\Big[t_0E+s_0C
-\sum_{i=i}^{N-1}\log(s_0+t_0\Lambda_i)\Big]\,;
\label{eq:gdefb}
\end{align}
the argument of the square root is positive, as seen from equation (\ref{eq:deldefb}).

The entropy of the microcanonical ensemble is
\begin{equation}
S(N,E,C)=\log\omega_N(E,C)=\Big[t_0E+s_0C-\sum_{i=i}^{N-1}\log(s_0+t_0\Lambda_i)\Big]
-\half \log(w_0w_2-w_1^2)+\hbox{const},
\end{equation}
where Boltzmann's constant has been set to unity. For fixed values of $s_0$, $t_0$,
$\Lambda_i$ the content of the square bracket grows linearly with $N$
while the second term grows only logarithmically. Thus as
$N\to\infty$ we may drop the second term, and
\begin{equation}
S(N,E,C)=t_0E+s_0C-\sum_{i=i}^{N-1}\log(s_0+t_0\Lambda_i)+\hbox{const}.
\end{equation}
The temperature $T$ is defined by
\begin{equation}
\frac{1}{T}=\left(\frac{\p S}{\p E}\right)_C=t_0 +\left(E\frac{\p t_0}{
    \p E} +C\frac{\p s_0}{\p E}\right)_C -\sum_{i=1}^{N-1}\frac{1}{
    s_0+t_0\Lambda_i}\left(\frac{\p s_0}{\p E}
    + \Lambda_i\frac{\p t_0}{\p E}\right)_C.
\end{equation}
Using equations (\ref{eq:eldef}), this and the analogous expression for $\p
S/\p C$ simplify to
\begin{equation}
\frac{1}{T}=\left(\frac{\p S}{\p E}\right)_C=t_0, \quad
\left(\frac{\p S}{\p C}\right)_E=s_0.
\end{equation}

We now investigate whether the solution to equations (\ref{eq:eldef}) exists
and is unique. By manipulating the identity
$N-1=\sum_{i=1}^{N-1}(s_0+t_0\Lambda_i)/(s_0+t_0\Lambda_i)$ it is
easy to see that
\begin{equation}
s_0C+t_0E=N-1.
\label{eq:sdef}
\end{equation}
With this result, equations (\ref{eq:eldef}) can be combined to give
\begin{equation}
\Gamma=\frac{E}{C}=h(N,C,\Gamma;t_0)\quad\hbox{where}\quad
h(N,C,\Gamma;t)\equiv \frac{\sum_{i=1}^{N-1} \Lambda_i[N-1+tC(\Lambda_i-\Gamma)]^{-1}}{
\sum_{i=1}^{N-1} [N-1+tC(\Lambda_i-\Gamma)]^{-1}}.
\label{eq:hdef}
\end{equation}
This non-linear equation can be solved for $t_0$ given $E$, $C$, and
$N$. Furthermore we have
\begin{equation}
\frac{\partial h}{\partial t}(N,C,\Gamma;t)=
-\frac{C(N-1)}{
  2\left[\sum_{i=1}^{N-1}[N-1+tC(\Lambda_i-\Gamma)]^{-1}\right]^2}
 \sum_{i,j=1}^{N-1}\frac{(\Lambda_i-\Lambda_j)^2}{
 [N-1+tC(\Lambda_i-\Gamma)]^2[N-1+tC(\Lambda_j-\Gamma)]^2},
\end{equation}
which is negative-definite, so $h$ decreases monotonically with $t$. Thus
there is at most one solution for $t_0$ for given $E$, $C$, and $N$. Using
equation (\ref{eq:sdef}) to eliminate $s_0$ from the constraints
(\ref{eq:lamlim}), and recalling that $\Lambda_{\rm
  min}\le\Gamma\le\Lambda_{\rm max}$ (see the discussion following eq.\
\ref{eq:omdefa}) we find that $t_0$ is restricted to the range
\begin{equation}
t_{\rm min}\equiv -\frac{N-1}{ C(\Lambda_{\rm max}-\Gamma)} < t_0 < t_{\rm
  max}\equiv \frac{N-1}{ C(\Gamma-\Lambda_{\rm min})}.
\end{equation}
Note that $t_{\rm min}<0$ and $t_{\rm max}>0$, and
\begin{equation}
h(t_{\rm min})=\Lambda_{\rm max}, \quad h(t_{\rm max})=\Lambda_{\rm min}.
\end{equation}
Thus $h(t)$ decreases monotonically from $\Lambda_{\rm max}$ to $\Lambda_{\rm
  min}$ as $t$ grows over its allowed range from $t_{\rm min}$ to $t_{\rm
  max}$. Since $\Gamma$ must lie between $\Lambda_{\rm max}$ to $\Lambda_{\rm
  min}$ for any initial condition, equation (\ref{eq:hdef}) always has a
unique solution for $t_0$ given $E$, $C$, and $N$, and equation
(\ref{eq:sdef}) then gives a unique solution for $s_0$.

Note that $h(N,C,\Gamma;0)=\sum_{i=1}^{N-1}\Lambda_i/N\equiv\Lambda_A$, the
arithmetic mean of the eigenvalues. Thus if $\Gamma>\Lambda_A$, $t_0<0$
(negative temperature) while if $\Gamma<\Lambda_A$ then $t_0>0$
(positive temperature).

The one-particle distribution function $f_1(P_1,Q_1)$ is defined so that
$f(P_1,Q_1)dP_1dQ_1$ is the probability that $P_1$, $Q_1$ lie in the phase-space
volume element $dP_1dQ_1$. We have
\begin{align}
f_1(P_1,Q_1)=&\frac{1}{\omega_N(E,C)}\int\prod_{i=2}^{N-1}dP_idQ_i\,
\delta\Big[E-\Lambda_1(P_1^2+Q_1^2)-\sum_{i=2}^{N-1}\Lambda_i(P_i^2+Q_i^2)\Big]
 \delta\Big[C-P_1^2-Q_1^2-\sum_{i=2}^{N-1}(P_i^2+Q_i^2)\Big] \notag\\
=&\frac{\omega_{N-1}\Big[E-\Lambda_1(P_1^2+Q_1^2),C-P_1^2-Q_1^2\Big]}{\omega_N(E,C)}.
\end{align}
Since $N\gg1$ the differences between the numerator and denominator in the
solutions of equations (\ref{eq:eldef}) for $s_0$ and $t_0$ are negligible, as
are the differences in $w_k$ defined by equation (\ref{eq:wwwdef}). With this
approximation and the use of equation (\ref{eq:gdefb}) we have
\begin{equation}
f_1(P_1,Q_1)=\frac{s_0+t_0\Lambda_1}{\pi}\exp\Big[-(s_0+t_0\Lambda_1)(P_1^2+Q_1^2)\Big].
\label{eq:gauss}
\end{equation}
Similar arguments show that the two-particle distribution function is the
product of one-particle functions,
$f_2(P_1,Q_1,P_2,Q_2)=f(P_1,Q_1)f(P_2,Q_2)$, etc.

The average energy in a single normal mode is
\begin{equation}
E_i\equiv \Lambda_i\langle
P_i^2+Q_i^2\rangle=\frac{\Lambda_i}{s_0+t_0\Lambda_i},
\end{equation}
where the last equality follows from (\ref{eq:gauss}). If $s_0=0$ each mode
contains the same energy; in other words, there is equipartition of energy or
the power spectrum is independent of the frequency $\Lambda_i$. We shall call
this `white noise' -- a small abuse of language since in most applications
white noise corresponds to constant energy per unit frequency rather than per
mode. If $s_0<0$ then $E_i$ grows as the frequency $\Lambda_i$ decreases (red
noise). If $s_0>0$ the energy per mode grows as the frequency grows (blue
noise). The condition $s_0=0$ corresponds to
$\Gamma=N/\sum_{i=1}^{N-1}\Lambda_i^{-1}\equiv\Lambda_H$, the harmonic mean of
the eigenvalues. Since $\Lambda_H<\Lambda_A$ (harmonic mean is less than the
arithmetic mean), we have three cases,
\begin{align}
\noalign{\mbox{(i) Positive temperature, red noise:}} \notag\\
\Lambda_{\rm min} < \Gamma < \Lambda_H, & \quad -\frac{(N-1)\Lambda_{\rm min}}{
  C(\Gamma-\Lambda_{\rm min})} < s_0 < 0, \quad \frac{(N-1)}{
  C(\Gamma-\Lambda_{\rm min})} > t_0 >
\frac{N-1}{C\Lambda_H}; \notag \\
\noalign{\mbox{(ii) Positive temperature, blue noise:}} \notag \\
\Lambda_H < \Gamma < \Lambda_A, & \quad 0 < s_0 < \frac{N-1}{C}, \quad
\frac{N-1}{C\Lambda_H} > t_0 > 0; \notag \\
\noalign{\mbox{(iii) Negative temperature, blue noise:}} \notag \\
\Lambda_A < \Gamma < \Lambda_{\rm max}, & \quad \frac{N-1}{C} < s_0 <
\frac{(N-1)\Lambda_{\rm max}}{ C(\Lambda_{\rm max}-\Gamma)}, \quad 0 > t_0
> -\frac{(N-1)}{ C(\Lambda_{\rm max}-\Gamma)}.
\end{align}

Microcanonical ensembles characterized by red noise tend to exhibit warps,
because there is more power in low-frequency, long-wavelength modes.

\subsection{Simulations of the microcanonical ensemble}

\label{sec:sim}

We want to generate realizations of a relaxed disc containing $N$ stars on
circular orbits, with masses and semi-major axes $m_i$, $a_i$,
$i=0,\ldots,N-1$. The disc is specified initially by its energy $E$ and
angular-momentum deficit $C$. The procedure is: (i) compute the matrices
$\bfssA$ (eq.\ \ref{eq:adef}) and $\bfssO$ (eq.\ \ref{eq:ortho}) and the
eigenvalues $\Lambda_i$ of $\bfssA$. (ii) Set $\Gamma=E/C$; if $\Gamma$ lies
outside the interval $[\Lambda_{\rm min},\Lambda_{\rm max}]$ spanned by the
non-zero eigenvalues $\Lambda_i$, $i=1,N-1$ then these values of $E$ and $C$
cannot be achieved by a disc with the given masses and semi-major axes.
Otherwise, find the unique solution $t_0$ to the non-linear equation
(\ref{eq:hdef}) and then find $s_0$ from the linear equation (\ref{eq:sdef}).
(iii) Set $P_0=Q_0=0$.  Choose $(P_i,Q_i)$, $i=1,\ldots,N-1$ at random from
the Gaussian distribution (\ref{eq:gauss}). (iv) Set
$p_i=\sum_{j=1}^{N-1}O_{ij}P_j$, $q_i=\sum_{j=1}^{N-1}O_{ij}Q_j$, then find
the inclinations $I_i$ and nodes $\Omega_i$ of the $N$ stars using equations
(\ref{eq:pqdef}). The resulting disc will have energy $E$ and angular-momentum
deficit $C$ that are within O$(N^{-1/2})$ of the assumed initial conditions
(although the differences can still be substantial if the energy or
angular-momentum deficit is dominated by a small number of modes).

Note that discs with different values of $E$ and $C$ but the same value of
$\Gamma=E/C$ have $t_0\propto 1/C$ (by eq.\ \ref{eq:hdef}) and $s_0\propto
1/C$ (by eq.\ \ref{eq:sdef}) so the distributions of $P_i$, $Q_i$, $p_i$,
$q_i$, and $I_i$ scale as $\surd C$ (by eq.\ \ref{eq:gauss}). Thus, apart from
this trivial scaling, the relaxed discs with a given distribution of stellar
masses and semi-major axes form a one-parameter family defined by the
frequency parameter $\Gamma$.

We generate microcanonical ensembles of discs having the same distribution of
stellar masses and semi-major axes as in the simulations described at the
start of \S\ref{sec:monte}. The eigenvalues $\Lambda_i$ in this disc span a
range of $10^7$ in magnitude (Fig.\ \ref{fig:Lambdahist}).  The largest
eigenvalues arise because the interactions of stars that happen to have
similar semi-major axes, $\Delta a/a\ll1$, are unrealistically strong,
essentially because the approximation that the Hamiltonian is quadratic is
only valid so long as the inclinations satisfy $I\ll\Delta a/a$. Thus the
calculations presented here may not accurately represent the thickness of the
equilibrium disc.

\begin{figure}
\centering
\includegraphics[width=84mm]{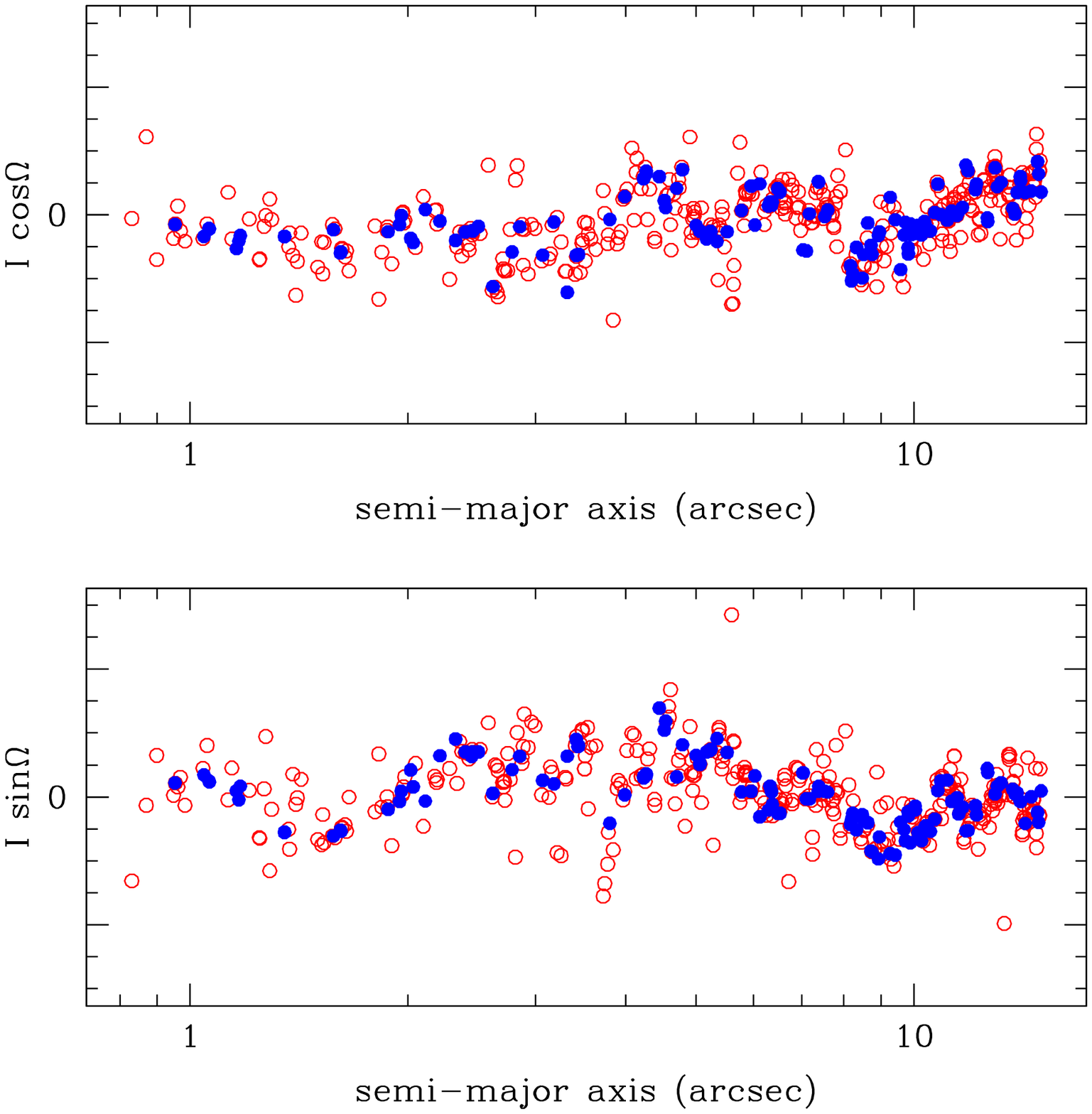}
\caption{A realization of the microcanonical ensemble. The upper and lower
  panels show $I\cos\Omega$ and $I\sin\Omega$ for a disc with 500 stars in the
  mass range 1 to $30\msun$; the mass function is given by equation
  (\ref{eq:imf}) with $\alpha=-0.45$ and the semi-major axis distribution is
  given by equation (\ref{eq:sd}) with $\gamma=-1.4$. The inner and outer
  semi-major axes are $1\arcsec$ and $15.5\arcsec$. Solid blue circles denote stars with
  mass $>20\msun$, corresponding roughly to stars that are visible in current
  surveys, and open red circles denote stars with smaller masses. The frequency parameter
  $\Gamma$ of eq.\ (\ref{eq:omdef}) equals $(10^5\yr)^{-1}$.
}
\label{fig:zero}
\end{figure}

A sample simulation with frequency parameter $\Gamma=(10^5\yr)^{-1}$ is
shown in Figure \ref{fig:zero}. The two plots show $I\cos\Omega$ and
$I\sin\Omega$ as a function of semi-major axis; the visible high-mass stars
($m>20\msun$) are marked with solid blue circles, and the low-mass stars are
marked by open red circles. The vertical axis is arbitrary since the disc
properties can be scaled to other energies, as described above.

As discussed above, the high-frequency modes are unphysical, because the
interactions between adjacent stars are unrealistically strong. However, most
of the structure visible in Figure \ref{fig:zero} arises from low-frequency
modes, which are largely independent of the strong interactions between close
neighbors. To illustrate this, we have experimented with softening the
interaction potential according to equation (\ref{eq:soft}). We find that
softenings as large as $\epsilon=0.003$ have very little effect on the
appearance of realizations of the disc.

One of the striking features of the disc shown in Figure \ref{fig:zero} is the
overall warp. To characterize the warp amplitude quantitatively, we first
define the inclination vectors $\bfI_i\equiv I_i(\cos\Omega_i,\sin\Omega_i)$
where $I_i$ and $\Omega_i$ are the inclination and node of star $i$. Then we
define the mean and dispersion of the $N_{\cal A}$ inclination vectors in some
semi-major axis range $\cal A$ by
\begin{equation}
\overline \bfI_{\cal A}\equiv \frac{1}{N_{\cal A}}\sum_{i\in\cal A} \bfI_i, \quad
\sigma_{\cal A}^2 \equiv \frac{1}{2(N_{\cal A}-1)}\sum_{i\in\cal
  A}(\bfI-\overline\bfI_{\cal A})^2;
\end{equation}
the factor of two in the second equation arises because $\sigma_{\cal A}^2$ is
the dispersion in one of the two components of the vector $\bfI$. In the
small-angle approximation within which we are working, the angle
between the mean orbit normals in regions $\cal A$ and $\cal A'$ is
\begin{equation}
\theta\equiv |\overline \bfI_{\cal A}-\overline \bfI_{\cal A'}|.
\end{equation}
A measure of the ratio of the warp to the disc thickness is then
$\theta/\sigma$ where $\sigma^2=\half(\sigma^2_{\cal A}+\sigma^2_{\cal A'})$.
This ratio is independent of the energy $E$ and angular-momentum deficit $C$
so long as the frequency parameter $\Gamma=E/C$ is fixed.

\begin{figure}
\centering
\includegraphics[width=84mm]{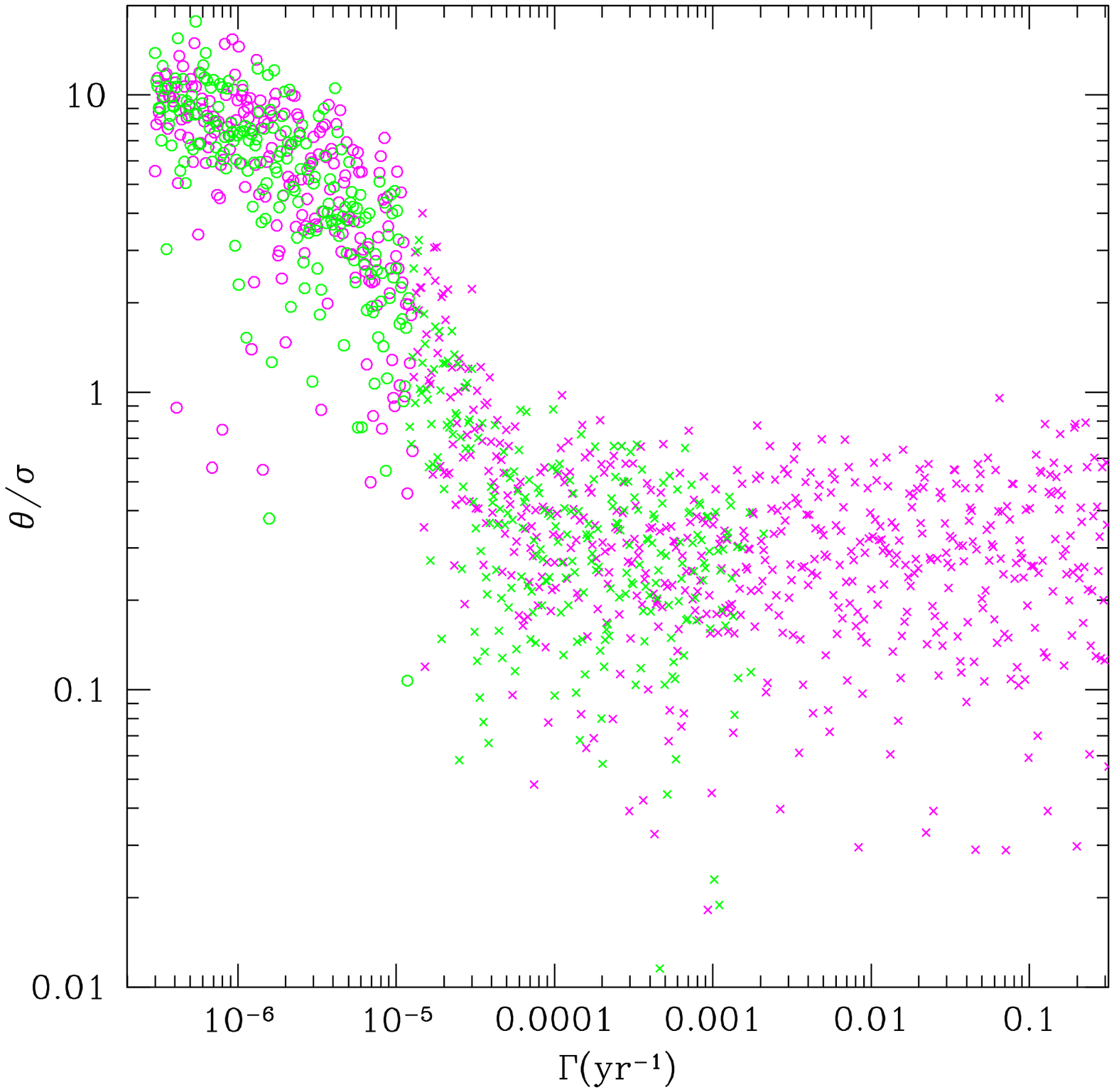}
\caption{Dimensionless disc warping $\theta/\sigma$ as a function of the
  frequency parameter $\Gamma$. The stellar semi-major axes and masses are
  chosen by the same procedure used to produce Fig.\ \ref{fig:zero}, with a
  different pseudorandom number seed for each point. The parameter $\theta$ is
  the angle between the mean inclination vectors in the inner and outer third
  of the disc, and $\sigma$ is a measure of the thickness of the disc (see
  text). Magenta and green points correspond to simulations with softening
  $\epsilon=0$ and 0.003, respectively. Open circles and crosses represent
  discs with red and blue noise ($\Gamma$ less or greater than the harmonic
  mean of the eigenvalues, $\Lambda_H$). Red discs exhibit large warps and
  blue discs do not.}
\label{fig:three}
\end{figure}

Warps are associated with large-scale, low-frequency normal modes, and
therefore are stronger in microcanonical ensembles dominated by red noise.
This is illustrated in Figure \ref{fig:three}, which shows the warp ratio
$\theta/\sigma$ for the visible stars ($m>20\msun$) in a set of disc
realizations with semi-major axes and stellar masses chosen as described at
the start of this subsection.  Magenta points are from unsoftened simulations
and green points are from simulations with softening $\epsilon=0.003$.  The
two regions compared, $\cal A$ and $\cal A'$, are the inner and outer third of
the disc stars.  Discs with red and blue noise ($\Gamma < \Lambda_H$ and
$\Gamma > \Lambda_H$, respectively) are marked by open circles and crosses.
The warp ratio $\theta/\sigma$ declines from $\sim 10$ for
$\Gamma\ll\Lambda_H$ (substantial warp) to $\sim 0.3$ for $\Gamma\gg\Lambda_H$
(negligible warp). The softened and unsoftened simulations exhibit the same
behavior, except that realizations with $\Gamma \gtrsim 0.003\hbox{ yr}^{-1}$
are not present in the softened simulations because the very high-frequency
normal modes involving two adjacent stars are suppressed by the softening. We
conclude that the microcanonical ensemble can exhibit significant large-scale
warps, depending on the initial conditions as described by $\Gamma$.

\end{document}